\documentclass[aps,prb]{revtex4}
\usepackage{graphicx}

\def \be{\begin{equation}}
\def \ee{\end{equation}}
\def \ba{\begin{array}}
\def \ea{\end{array}}
\def \bea{\begin{eqnarray}}
\def \eea{\end{eqnarray}}
\def \nn{\nonumber}

\def \half{{1\over 2}}

\def \bQ{{\bf Q}}

\def \bk{{\bf k}}

\def \e{{\epsilon}}

\def \a{{\alpha}}

\def \b{{\beta}}
\def \g{{\gamma}}

\def \d{{\delta}}
\def \w{{\omega}}

\def \f{{\varphi}}
\def \x{{\chi}}

\def \yd{^\dagger}
\def \av#1{{\langle#1\rangle}}

\begin{document}

\title{Competition between Triplet
Superconductivity and Antiferromagnetism
in Quasi One-Dimensional Electron Systems}

\author{Daniel Podolsky,  Ehud Altman, Timofey Rostunov, and Eugene Demler}
\address{Department of Physics, Harvard University, Cambridge MA 02138}
\date{\today}

\begin{abstract}
We investigate the competition between antiferromagnetism and
triplet superconductivity in quasi one-dimensional electron
systems.  We show that the two order parameters can be unified
using a SO(4) symmetry and demonstrate the existence of such
symmetry in one dimensional Luttinger liquids of interacting
electrons.  We argue that approximate SO(4) symmetry remains valid
even when interchain hopping is strong enough to turn the system
into a strongly anisotropic Fermi liquid. For unitary triplet
superconductors SO(4) symmetry requires a first order transition
between antiferromagnetic and superconducting phases. Analysis of
thermal fluctuations shows that the transition between the normal
and the superconducting phases is weakly first order, and the
normal to antiferromagnet phase boundary has a tricritical point,
with the transition being first order in the vicinity of the
superconducting phase. We propose that this phase diagram explains
coexistence regions between the superconducting and the
antiferromagnetic phases, and between the antiferromagnetic and
the normal phases observed in (TMTSF)$_2$PF$_6$. For non-unitary
triplet superconductors the SO(4) symmetry predicts the existence
of a mixed phase of antiferromagnetism and superconductivity.  We
discuss experimental tests of the SO(4) symmetry  in neutron
scattering and tunneling experiments.

\end{abstract}

\maketitle



\section{Introduction}
\label{SectionIntro}

Quasi one-dimensional compounds can display a rich variety of
phases, including spin-Peierls, charge density wave, spin density
wave, and superconducting orders\cite{Ishiguro1998,Jerome1994,
Coleman1973,Bechgaard1980,Maaroufi1985}. Due to the large
anisotropy in their crystal structure, these materials are often
modelled as a collection of weakly coupled Luttinger liquids.  The
wealth of phases seen in these compounds is then attributed to the
intrinsic instability of one dimensional electron systems towards
the formation of quasi long range order\cite{Barisic1981}. As
temperature is lowered, correlations along individual chains grow,
until the coupling between chains stabilize true long range order.
In the current paper, we follow this approach to study the
interplay between triplet superconductivity (TSC) and
antiferromagnetism (AF) in quasi one-dimensional electron systems.
The starting point of our discussion is an observation that, for
weak umklapp scattering, one-dimensional Luttinger liquids at
half-filling have SO(4) symmetry at the boundary between AF and
TSC phases.  Near this boundary, the two order parameters can be
unified using SO(4) symmetry, leading to strong constrains on the
topology of the phase diagram and on the spectrum of low energy
collective excitations.

Our analysis is motivated by quasi one-dimensional Bechgaard salts
(TMTSF)$_2$X, and their sulphurated counterparts (TMTTF)$_2$X. The
most well studied material from this family (TMTSF)$_2$PF$_6$ is
an antiferromagnetic insulator at ambient pressure and becomes a
superconductor at high
pressure\cite{Jerome1980,Andres1980,Takahashi1989,Vuletic2002,Kornilov2003}.
The symmetry of the superconducting order parameter in
(TMTSF)$_2$PF$_6$ is not yet fully established\cite{Dressel1999},
but there is strong evidence that electron pairing is spin
triplet: the superconducting T$_c$ is strongly suppressed by
disorder
\cite{Choi1984,Choi1982,Tomic1983,Coulon1982,Abrikosov1983};
critical magnetic field $H_{c2}$ in the interchain direction
exceeds the paramagnetic limit \cite{Lee1997,Gorkov1985}; the
electron spin susceptibility, obtained from the Knight shift
measurements, does not decrease below Tc \cite{Lee2000}. In
another material from this family, (TMTSF)$_2$ClO$_4$,
superconductivity is stable at ambient pressure and also shows
signatures of triplet
pairing\cite{Takigawa1987,Ha2003,Joo2004,Oh2004}.  Insulator to
superconductor transition as a function of pressure has also been
found for (TMTSF)$_2$AsF$_6$\cite{Brusetti1982} and
(TMTTF)$_2$PF$_6$\cite{Jaccard2000}.

There are two aspects of the SO(4) symmetry between
antiferromagnetism and triplet superconductivity that we address
in this paper.
\begin{trivlist}
\item {\it Classical SO(4) Symmetry}.~ We consider the possible
emergence of the classical (static) symmetry at a finite
temperature critical point. We introduce a Ginzburg-Landau (GL)
free energy to describe the interaction between the AF and TSC
orders, and we study the effects of thermal fluctuations through a
large $N$ expansion and renormalization group (RG) analyses in
$d=4-\epsilon$ and $d=2+\epsilon$ dimensions.  For a unitary TSC,
which we argue to describe Bechgaard salts, we find a first order
transition between AF and TSC phases, a first order transition
between AF and normal phases ending in a tricritical point, and a
weakly first order transition between TSC and normal phases.  For
a non-unitary TSC we find a mixed phase in which AF and TSC orders
are present simultaneously. We argue that the system is close to
having an SO(4) symmetric tetracritical point, but there is a
narrow line of direct first order transitions between the normal
and the mixed phase. (For a detailed discussion of the distinction
between unitary and non-unitary TSC, see Sec.~\ref{incommensurateSection}.)

\item {\it Quantum SO(4) Symmetry}.~
We introduce a quantum SO(4) rotor model which encapsulates key
features of the competition between AF and TSC orders.  We use
this model to study collective excitations in the system in
various phases. We argue that the $\Theta$-excitation, which gives
one of the generators of the SO(4) algebra, should give rise to a
sharp resonance in spin polarized neutron scattering in the TSC
phase. We further predict that in the case of a unitary TSC
the energy of the $\Theta$-resonance should decrease to nearly zero at the phase
boundary with the AF phase.  Such mode softening is not expected
generally near a first order transition and would be a unique
signature of the enhanced symmetry at the transition point.

\end{trivlist}

Bechgaard salts belong to a class of strongly correlated electron
systems displaying proximity of a superconducting state to some
kind of magnetically ordered insulating state. Other examples
include the high Tc cuprates\cite{Maple1998}, heavy fermion
superconductors\cite{Mathur1998,Kitaoka2001}, and in most cases
the superconducting (SC) order parameter is spin singlet ($s$ or
$d$ wave) and the insulating state has antiferromagnetic or spin
density wave order.  Symmetry principles have been introduced to
study the competition of order parameters in some of these
systems. In S.C. Zhang's SO(5) theory of high T$_c$
superconductivity\cite{Zhang1997}, antiferromagnetism and $d$-wave
superconductivity are treated as components of a five dimensional
order parameter.  In addition to the generators of the usual
charge SO(2) and spin SO(3) symmetries, new $\pi$-operators are
introduced, which rotate superconductivity and antiferromagnetism
into each other. A combination of analytical approximations and
numerical results can be used to argue an approximate SO(5) theory
of a class of two dimensional lattice models, such as the Hubbard
and the $t$-$J$ model \cite{Meixner1997,Eder1999}.  The SO(5)
symmetry has also been used to discuss quasi two-dimensional
organic $\kappa$-BEDT-TTF salts\cite{Murakami2000}. The
unification approach based on higher symmetries has been
generalized to several other types of competing states. SO(5) and
SO(8) symmetries have been used to classify possible many-body
ground states in electronic ladders \cite{Scalapino1998,Lin1998}.
SO(6) symmetry has been introduced to discuss competing striped
phases and superconductivity in the cuprates\cite{Markiewicz1998}
SO(4) symmetry has been used to combine $s$-wave superconductivity
and charge density wave orders in the negative $U$ Hubbard
model\cite{Yang1989,Zhang1990}, as well as $d$-wave
superconductivity and $d$-density wave phases
\cite{PLee1998,Nayak2000}. It was also suggested that the SO(5)
algebra can be used to combine ferromagnetism and triplet
superconductivity in quasi two-dimensional Sr$_2$RuO$_4$
\cite{Murakami1999}, although the existence of microscopic models
with such symmetry has not been demonstrated.

There are several reasons why Bechgaard salts, and
(TMTSF)$_2$PF$_6$ in particular, are promising candidates for
experimental observation of the emergence of high symmetry from
the competition of two orders. The insulator to superconductor
transition in these materials is tuned by pressure, so the entire
phase diagram can be explored in a single sample.  This compares
favorably to the cuprate superconductors, where the AF/SC
transition appears as a function of doping and different samples
are required to investigate various regimes. Another important
advantage of Bechgaard salts is that they may be well described by
a microscopic Luttinger liquid Hamiltonian, for which we can
demonstrate the existence of SO(4) symmetry using standard
bosonization analysis. This is in contrast to the high Tc
cuprates, in which approximations need to be made in order to even
define generators of the SO(5) symmetry
\cite{Zhang1997,Henley1998,Rabello1998}. A related issue is the
question of quasiparticles in the AF insulating state and in the
$d$-wave superconducting phase. In the former case the
quasiparticle spectrum is fully gapped while in the latter case
there are nodal quasiparticles.  It is not presently known how
this difference affects a quantum SO(5) symmetry for collective
bosonic degrees of freedom.  An advantage of the SO(4) symmetry in
(TMTSF)$_2$PF$_6$ is that quasiparticles are fully gapped in both
the superconducting and the insulating phases.


Recent neutron scattering experiments demonstrated the existence
of strong AF fluctuations in a triplet superconductor
$Sr_2RuO_4$\cite{Mackenzie2003,Braden2003,Friedt2003}. This
material is not quasi one-dimensional, but it has nested pieces of
the Fermi surface (see e.g. Ref. \onlinecite{Mackenzie2003}).
Thus, we expect that this material may also show some qualitative
features of the competition between AF and TSC discussed in this
paper.

We note that our approach is phenomenological in nature, since we
do not attempt to obtain Luttinger parameters starting from
microscopic considerations.  Instead, we observe that Bechgaard
salts remain strongly anisotropic even close to the AF/TSC phase
boundary. Hence, we argue that the Luttinger parameter should be
such that individual 1d chains should be in the vicinity of such
phase transition.  By starting with the Luttinger Hamiltonian, we
derive the SO(4) symmetry as its immediate consequence. We note,
however, that the Luttinger liquid physics is not a necessary
requirement for observing SO(4) symmetry near the AF/TSC phase
boundary. Several groups have argued that near the TSC phase of
Bechgaard salts, the interchain tunneling is sufficient to
suppress Luttinger liquid behavior in favor of a strongly
anisotropic Fermi liquid\cite{Bourbonnais1999,Dressel2004}.  We
will argue below that an approximate classical SO(4) symmetry will
be present near the AF/TSC boundary even if the ordered phases
arise from a Fermi liquid state, although we still rely on the
assumption that interchain hopping of electrons is much smaller
than intrachain hopping (this condition is satisfied for Bechgaard
salts, see Sec.~\ref{SectionInterchain}). Similarly, we expect
that the $\Theta$ resonance will be present even in a strongly
anisotropic Fermi liquid, whose observation will verify the
approximate quantum SO(4) symmetry. In this paper, for
concreteness, we will concentrate on the case where the ordered
phases emerge from Luttinger liquid behavior on individual chains.


This paper is organized as follows. In section
\ref{LuttingerSection} we discuss the Luttinger liquid model for
interacting electrons in one dimension.  For incommensurate band
filling, we show that along the transition line between the TSC
and the SDW phases, this model has SO(3)$\times$SO(4) symmetry. At
half-filling we argue that, for weak umklapp, this symmetry is
reduced to SO(4) symmetry.  For quasi one-dimensional systems such
as Bechgaard salts we argue that this SO(4) symmetry provides a
unified description of AF and TSC orders. In section
\ref{SectionGLDiscussion} we discuss a general GL free energy for
the interplay between magnetism and triplet superconductivity at
finite temperatures, and present mean field diagrams for these
orders. In section \ref{SectionThermal} we analyze thermal
fluctuations using $4-\epsilon$ RG analysis and demonstrate the
absence of stable fixed points, which could control multicritical
points in the phase diagram. In section \ref{SectionUnitaryTSC} we
analyze the case of unitary TSC competing with AF by extending the
spin SO(3) group to an SO(N) algebra and using large $N$ analysis.
In section \ref{SectionNonUnitary} we investigate the interplay of
non unitary TSC and AF using large N approach and RG analysis for
$N>3$ in $4-\epsilon$ and $2+\epsilon$ dimensions. We also discuss
a physically relevant case of $N=3$. In Section
\ref{QuantumSection} we introduce an effective SO(4) quantum rotor
model that condenses the essential features of the competition
between the two phases. We use this model to study collective
excitations in various phases.  In section~\ref{SectionInterchain}
we discuss SO(4) symmetry in highly anisotropic Fermi liquids.  In
section \ref{SectionSignatures} we review experimental
implications of the SO(4) symmetry for Bechgaard salts.  Finally,
in section \ref{SectionSummary} we summarize our results.

\section{Microscopic origin of the symmetry}
\label{LuttingerSection}

\subsection{SO(3)$\times$SO(4) symmetry at incommensurate filling}

\begin{figure}
\includegraphics[width=7.5cm]{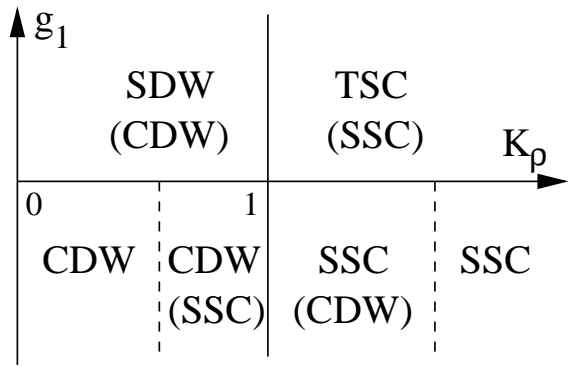}
\caption{Phase diagram for a one dimensional system of interacting
spin-1/2 fermions [\onlinecite{Giamarchi1989}]. Here
$K_\rho^2=(2\pi v_f +2 g_4 +g_1 -2 g_2)/(2\pi v_f +2 g_4 -g_1 +2
g_2)$. SDW and CDW correspond to spin and charge density wave
states, SS and TS to singlet and triplet superconducting phases.
\label{LuttingerPhaseDiagram}}
\end{figure}

Consider a one dimensional electron gas
with the Hamiltonian
\begin{eqnarray}
{\cal H}&=&{\cal H}_0 + {\cal H}_1 + {\cal H}_2 + {\cal H}_4
\nonumber\\
{\cal H}_0 &=&
\sum_{rks} (\epsilon_{r,ks} -\mu) a_{r,ks}^\dagger a_{r,ks}
\nonumber\\
{\cal H}_1 &=& \frac{g_1}{L}\sum
a_{+,ks}^\dagger a_{-,pt}^\dagger
a_{+,p+qt}a_{-,k-qs}
\nonumber\\
{\cal H}_2 &=& \frac{g_2}{L}\sum
a_{+,k+qs}^\dagger a_{-,p-qt}^\dagger
a_{-,pt}a_{+,ks}
\nonumber\\
{\cal H}_4 &=& \frac{g_4}{L}\sum
a_{+,k+qs}^\dagger a_{+,p-qt}^\dagger
a_{+,pt}a_{+,ks}
+ \frac{g_4}{L}\sum
a_{-,k+qs}^\dagger a_{-,p-qt}^\dagger
a_{-,pt}a_{-,ks}
\label{Luttinger_Hamiltonian}
\end{eqnarray}
Here $a_{\pm,ks}^\dagger$ create right/left moving electrons with
momenta $\pm k_f+k$ and spin $s$, and we assume linearized
dispersion of electrons $ \epsilon_{r,ks} -\mu = r v_f k $.  In
the Hamiltonian (\ref{Luttinger_Hamiltonian}) the interaction term
$g_1$ describes backward scattering and terms $g_2$ and $g_4$
describe forward scattering.  For now, we assume that the system
has incommensurate filling, so that umklapp processes are not
allowed. The phase diagram for this system obtained from the
renormalization group analysis has been discussed extensively
before (see e.g. Refs. \onlinecite{Solyom1979} and
\onlinecite{Giamarchi1989}) and is shown in Fig.
\ref{LuttingerPhaseDiagram}.
For the current discussion, we concentrate on the
region of the phase diagram near the transition line
between the TSC and the SDW phases at $K_\rho=1$, i.e.
\begin{eqnarray}
g_1=2g_2
\label{LuttingerSO4Condition}
\end{eqnarray}
We demonstrate that on this line the system has an
SO(3)$\times$SO(4) symmetry that unifies order parameters of the
two phases.

The total spin operators are defined as
\begin{eqnarray}
S_\a= \frac{1}{2} \sum_{r,kss'}a_{r,ks}^\dagger \sigma^\a_{ss'},
a_{r,ks'}
\end{eqnarray}
where $\sigma^\a_{ss'}$ are the usual Pauli matrices. These operators
form a spin SO(3) algebra
\begin{eqnarray}
\left[ S_\a, S_\b \right]= i \epsilon^{\a\b\gamma} S_\gamma
\label{SpinLuttinger}
\end{eqnarray}
We can also combine the charge operators for right and left movers ($r=\pm$),
\begin{eqnarray}
Q_{r}= \frac{1}{2} \sum_{ks} \left( a_{r,ks}^\dagger a_{r,ks}
-\frac{1}{2} \right) \label{chiralQ}
\end{eqnarray}
and the operators
\begin{eqnarray}
\Theta_r^\dagger=r\sum_k a_{r,k\uparrow}^\dagger
a_{r,-k\downarrow}^\dagger
\label{ThetaOperator}
\end{eqnarray}
to form two separate isospin SO(3) algebras
\begin{eqnarray}
J_x^r &=& \frac{1}{2}( \Theta_r^\dagger + \Theta_r)
\nonumber\\
J_y^r &=& \frac{1}{2i}( \Theta_r^\dagger - \Theta_r)
\nonumber\\
J_z^r &=& Q_r
\nonumber\\
\left[ J_a^r, J_b^{r'} \right] &=& i \delta_{r,r'}\epsilon^{abc} J_c^r
\label{IsospinLuttinger}
\end{eqnarray}
The total isospin group is therefore SO(4)$_{\rm
isospin}=$SO(3)$_R\times$SO(3)$_L$. Note that, since spin and
isospin operators commute, $[S_\a,J_b^r]=0$, they jointly define a
closed SO(3)$_{\rm spin}\times$SO(4)$_{\rm isospin}$ algebra.

The total spin, $S_\a$, and the total charge, $Q_++Q_-$,  always
commute with the Hamiltonian (\ref{Luttinger_Hamiltonian}). In
addition, due to the absence of umklapp at incommensurate filling,
$Q_+$ and $Q_-$ are conserved separately.  As shown in Appendix
\ref{Bosonization} using bosonization, when the condition
(\ref{LuttingerSO4Condition}) is satisfied, the $\Theta_r$
operators also commute with the Hamiltonian.  Hence, the system
has full SO(3)$\times$SO(4) symmetry at the phase boundary between
TSC and SDW phases.  We emphasize that the SO(3)$\times$SO(4)
symmetry of Luttinger liquids at the SDW/TSC boundary is generic
and does not require fine tuning of the parameters. SO(4)$_{\rm
isospin}$ invariance has been discussed in quasi one-dimensional
systems with highly anisotropic spin
interactions\cite{Rozhkov2002,Carr2002}.   The $\Theta_r$ operators
in (\ref{ThetaOperator}) are reminiscent of the $\eta$ operators
introduced by C.N. Yang to study the Hubbard model\cite{Yang1989,Zhang1990},
but we will show in Sec. \ref{SectionHalfFillingGL} that the two
sets of operators define different symmetry groups
and apply to different systems.

Spin density wave order away from half-filling is described by a
complex vector order parameter,
\be \Phi_\a = \sum_{kss'}
a^\dagger_{+,ks} \sigma^\a_{ss'} a_{-,ks'},
\label{SDWorder} \ee
For quasi one-dimensional systems, the band structure restricts
the orbital component of the triplet superconducting order to be
$\vec{\Psi}(\vec{p})\propto p_x$, where $x$ is the direction
parallel to the chains. Thus, the TSC order parameter is also
described by a complex vector,
\begin{eqnarray}
\Psi_\a^\dagger = \frac{1}{i} \sum_{kss'} a^\dagger_{+,ks}
(\sigma^\a \sigma_2)_{ss'} a^\dagger_{-,-ks'}
\end{eqnarray}
The factor of $-i$ is introduced for convenience,
$-i\sigma_2\equiv\left(\begin{array}{c c} 0 & -1 \\ 1 & 0 \end{array}\right).$
The four vector order parameters Re$\vec{\Phi}$, Im$\vec{\Phi}$, Re$\vec{\Psi}$,
and Im$\vec{\Psi}$ can be combined into a 4$\times$3 matrix,
\begin{eqnarray}
\hat{P} &=& \left(\begin{array}{cccc}
({\rm Re} \vec{\Psi})_x & ({\rm Im} \vec{\Psi})_x & ({\rm Re} \vec{\Phi})_x & ({\rm Im} \vec{\Phi})_x\\
({\rm Re} \vec{\Psi})_y & ({\rm Im} \vec{\Psi})_y & ({\rm Re} \vec{\Phi})_y & ({\rm Im} \vec{\Phi})_y\\
({\rm Re} \vec{\Psi})_z & ({\rm Im} \vec{\Psi})_z & ({\rm Re} \vec{\Phi})_z & ({\rm Im} \vec{\Phi})_z\\
\end{array}
\right)
\label{Pmatrix}
\end{eqnarray}
Each column of $\hat{P}$ transforms independently as a vector
under the action of the spin group,
\begin{eqnarray}
\left[{S_\alpha,P_{b\beta}} \right] &=&
i\epsilon^{\alpha\beta\gamma}P_{b\gamma}
\end{eqnarray}
The action of the isospin group on $\hat{P}$ is easiest to understand in terms of the operators
$I_a=J_a^++J_a^-$ and $\Lambda_a=J_a^+-J_a^-$.  For a fixed row of $\hat{P}$,
the action of the isospin generators
in the basis
$\left({\rm Re} \Psi_\a , {\rm Im} \Psi_\a , {\rm Re} \Phi_\a  ,  {\rm Im} \Phi_\a \right)$
is represented by
\begin{eqnarray}
\begin{array}{cc}
I_x=\left(\begin{array}{cccc}
0&0&0&0\\
0&0&-i&0\\
0&i&0&0\\
0&0&0&0
\end{array}\right)
& \Lambda_x=\left(\begin{array}{cccc}
0&0&0&-i\\
0&0&0&0\\
0&0&0&0\\
i&0&0&0\end{array}\right) \\
I_y=\left(\begin{array}{cccc}
0&0&-i&0\\
0&0&0&0\\
i&0&0&0\\
0&0&0&0\end{array}\right)
& \Lambda_y =\left(\begin{array}{cccc}
0&0&0&0\\
0&0&0&i\\
0&0&0&0\\
0&-i&0&0\end{array}\right)\\
I_z=\left(\begin{array}{cccc}
0&i&0&0\\
-i&0&0&0\\
0&0&0&0\\
0&0&0&0\end{array}\right)
& \Lambda_z =\left(\begin{array}{cccc}
0&0&0&0\\
0&0&0&0\\
0&0&0&i\\
0&0&-i&0\end{array}\right)
\end{array}
\label{Ggen}
\end{eqnarray}
Once the fourth component is identified as the ``time-like"
direction, this is the (Euclidean) Lorentz group, with the $I_a$
acting as rotations and the $\Lambda_a$ acting as boosts.  Hence, the rows of
order parameter $P_{\bar{a}\a}$ transform in the vector representation of
the SO(4)$_{\rm isospin}$ group.

\subsection{SO(4) symmetry at half filling}

In Bechgaard salts (TMTSF)$_2$X, three out of every four
conduction states are occupied.  At quarter filling, umklapp
processes involving interactions of four electrons are allowed.
Such interactions are weak, and furthermore are irrelevant in the
RG sense for $K_\rho>1/4$ (we remind the readers that we are
interested in the regime near the SDW/TSC boundary, where
$K_\rho\approx 1$)\cite{Giamarchi1997}. On the other hand, due to
structural dimerization in Bechgaard salts\cite{Thorup}, a gap
splits the conduction band into a completely filled lower band and
a half-filled upper band. Hence, Bechgaard salts are half-filled
systems. At half filling, the Hamiltonian
(\ref{Luttinger_Hamiltonian}) must be modified to include
two-electron umklapp scattering processes,
\begin{eqnarray}
{\cal H}_3 &=& \frac{g_3}{2L}\sum
a_{+,k+qs}^\dagger a_{+,p-qt}^\dagger
a_{-,pt}a_{-,ks}
+ \frac{g_3}{2L}\sum
a_{-,k+qs}^\dagger a_{-,p-qt}^\dagger
a_{+,pt}a_{+,ks}
\label{umklapp}.
\end{eqnarray}
Analysis of the phase diagram of Luttinger liquids at half-filling
reveals that there is still a direct transition between AF and TSC
orders at $K_\rho=1$, although this condition now
corresponds\cite{Jerome1994} to $g_1-2g_2=|g_3|$. The umklapp term
allows scattering of two right moving electrons into two left
moving ones, and vice versa. Thus, it does not commute with the
operator $\Lambda_z=Q_+-Q_-$, which leads to breaking of the
SO(3)$\times$SO(4) symmetry.  To understand the nature of this
symmetry breaking, it is useful to rewrite (\ref{umklapp}) in the
form, \be {\cal H}_3 = \frac{g_3}{2L}\sum_q \left( {\rm
Re}\hat{\vec{\Phi}}(q)\cdot {\rm Re}\hat{\vec{\Phi}}(-q)- {\rm
Im}\hat{\vec{\Phi}}(q)\cdot {\rm Im}\hat{\vec{\Phi}}(-q) \right),
\label{umkSDW} \ee where $\hat{\vec{\Phi}}(\vec{q})$ is the SDW
order parameter at center of mass momentum $2k_f+\vec{q}$, \be
\hat{\vec{\Phi}}(\vec{q})=\sum_{kss'}a^\dagger_{+,ks}\vec{\sigma}_{ss'}a_{-,k-qs'}.
\ee Equation (\ref{umkSDW}) shows explicitly that umklapp tends to
pin the phase of the SDW order parameter at either 0 or $\pi$,
depending on the sign of $g_3$. This is in agreement with the
observation that period two antiferromagnetic order can be
described by a single real N\'eel vector.

We will show in Section \ref{SectionHalfFillingGL} that, whereas
the Ginzburg-Landau free energy is no longer SO(3)$_{\rm
spin}\times$SO(4)$_{\rm isospin}$ symmetric at half-filling, to
linear order in $g_3$ it maintains an SO(4)=SO(3)$_{\rm
spin}\times$SO(3)$_{\rm isospin}$ symmetry.  The unbroken part of
the isospin group, SO(3)$_{\rm isospin}$, is the diagonal subgroup
of SO(3)$_R\times$SO(3)$_L$, which is generated by the three $I_a$
operators, $I_x = \frac{1}{2}( \Theta^\dagger + \Theta),$ $I_y =
\frac{1}{2i}( \Theta^\dagger - \Theta),$ $I_z = Q$, where
\begin{eqnarray}
Q&=& \frac{1}{2} \sum_{ks} \left( a_{+,ks}^\dagger a_{+,ks}
+ a_{-,ks}^\dagger a_{-,ks} -1 \right) \nonumber\\
\Theta^\dagger&=&\sum_k \left( a_{+,k\uparrow}^\dagger
a_{+,-k\downarrow}^\dagger -  a_{-,k\uparrow}^\dagger
a_{-,-k\downarrow}^\dagger \right).
\label{quantumGenerators}
\end{eqnarray}
Without loss of generality we consider the case $g_3<0$,
where the order parameter for antiferromagnetism
is given by the real part of $\vec{\Phi}$,
\begin{eqnarray}
N_\a = \frac{1}{2}
\sum_{kss'} \left( a^\dagger_{+,ks} \sigma^\a_{ss'} a_{-,ks'}
+ a^\dagger_{-,ks} \sigma^\a_{ss'} a_{+,ks'} \right)
\label{AFOrderParamater}
\end{eqnarray}
It is easy to verify that $\{ \vec{N}, {\rm Re} \vec{\Psi},
{\rm Im} \vec{\Psi}\}$
transform as vectors under both spin and isospin SO(3)
symmetries. We define
\begin{eqnarray}
\hat{Q} &=& \left(\begin{array}{ccc}
({\rm Re} \vec{\Psi})_x & ({\rm Im} \vec{\Psi})_x & N_x \\
({\rm Re} \vec{\Psi})_y & ({\rm Im} \vec{\Psi})_y & N_y \\
({\rm Re} \vec{\Psi})_z & ({\rm Im} \vec{\Psi})_z & N_z
\end{array}\right)
\label{Qmatrix}
\end{eqnarray}
$\hat{Q}$ transforms as a vector under both SO(3)
algebras
\begin{eqnarray}
\left[ {I_a,Q_{b\beta}} \right]&=& i\epsilon^{abc}Q_{c\beta}
\nonumber\\
\left[{S_\alpha,Q_{b\beta}} \right] &=& i\epsilon^{\alpha\beta\gamma}Q_{b\gamma}
\label{OrderParameterLuttinger}
\end{eqnarray}
so it describes an order parameter that transforms
as a (1,1) representation of the SO(4) algebra.  Since we are mostly interested
in applying our results to Bechgaard salts, we focus mostly on this SO(4) symmetry
in Ref. \onlinecite{Podolsky2004}, as well as on the remainder of
this paper.

Unlike the SO(3)$\times$SO(4) symmetry discussed in the
incommensurate case, the SO(4) symmetry at half-filling is not a
rigorous symmetry of the system.  The generators of this group do
not commute with the Hamiltonian of the system exactly. However,
the main emphasis of our work is to understand the finite
temperature phase diagram of (TMTSF)$_2$PF$_6$.  This is obtained
from the classical GL free energy, which at the AF/TSC phase
boundary has SO(4) symmetry if we retain umklapp processes to
linear order in $g_3$ (see discussion in Section
\ref{SectionHalfFillingGL}). In addition, with regards to quantum
properties, SO(4) symmetry is a good starting point to study the
collective modes of the system when $g_3$ is small.  The latter
assumption is well justified for (TMTSF)$_2$PF$_6$, since the
observed dimerization in this material is less than
1$\%$\cite{Thorup}. For small $g_3$, modes found assuming SO(4)
symmetry will have a finite overlap with the actual excitations of
the system. In particular, the quantum numbers of the $\Theta$
mode discussed in Section \ref{QuantumSection}, including charge
two and center of mass momentum $2k_f$, are not affected by
umklapp. These properties determine which experimental probes
couple to $\Theta$. We must keep in mind, however, that the
explicit breaking of SO(4) due to higher order corrections in
$g_3$, and also due to interchain coupling, may give a small
energy gap and finite broadening to $\Theta$, even at the AF/TSC
phase boundary.  We also point out that, from the point of view of
$\Theta$ excitations, the difference between SO(3)$\times$SO(4)
and SO(4) symmetries corresponds to the question whether $+2k_f$
and $-2k_f$ excitations are the same (SO(4) symmetry at half
filling) or different (SO(3)$\times$SO(4) symmetry away from half
filling).

Thus far in the analysis we have ignored spin-orbit effects.
Microwave absorption experiments in (TMTSF)$_2$AsF$_6$ measured
\cite{Torrance1982} the anisotropy in the exchange couplings to be
$10^{-6}$. This ultimately determines the preferred axes for the
N\'eel vector ${\vec{N}}$ (along the b-axis of the crystal
\cite{Mortensen1982}) and the spin component of the TSC order
${\vec{\Psi}}$ (along either the a- or c-axis \cite{Lee1997}).
However, we do not expect such tiny anisotropy to play a
significant role in determining the competition between AF and TSC
phases.  We also point out that NMR experiments in
(TMTSF)$_2$PF$_6$ find a divergence of $T_1^{-1}$ at the N\'eel
temperature that is well-described by the O(3) isotropic
Heisenberg model \cite{Bourbonnais1986}.  Thus, even for critical
fluctuations of the AF order parameter, spin anisotropy coming
from spin-orbit coupling is unobservably small.

Before concluding this section we point out that the isospin
algebra defined by equations (\ref{chiralQ}),
(\ref{ThetaOperator}), and (\ref{IsospinLuttinger}) can be also
used to relate charge density wave order and singlet
superconductivity in quasi one-dimensional electron
systems\cite{Efetov1981}. This is relevant for the lower half of
Fig. \ref{LuttingerPhaseDiagram}.

\section{Ginzburg-Landau Free Energy}
\label{SectionGLDiscussion}


The main goal of this section is to investigate consequences of
the SO(4) symmetry for the true finite temperature phase
transitions, when we need to consider three dimensional
fluctuations of the order parameter.  One may be concerned that by
introducing interchain couplings, we will immediately destroy the
SO(4) symmetry.  As we will show in
section~\ref{SectionInterchain}, even in the case where the
interchain coupling $t_b$ is large enough to make the system into
a highly anisotropic Fermi liquid, approximate SO(4) symmetry
prevails in the Ginzburg-Landau (GL) free energy (see e.g. Eq.
(\ref{NearlySO4GLenergy}). This feature of the AF/TSC GL free
energy implies that our analysis of the phase diagram, based on
classical SO(4) symmetry, is valid even when the normal state is
described by a highly anisotropic Fermi liquid rather than a
collection of weakly coupled Luttinger liquids. A review of the
normal state properties of organic superconductors at low magnetic
fields is given in Ref.~\onlinecite{Bourbonnais1999}, and low
temperature transport properties have been reported recently in
Ref.~\onlinecite{Dressel2004}.

\begin{figure}
\includegraphics[width=7.5cm]{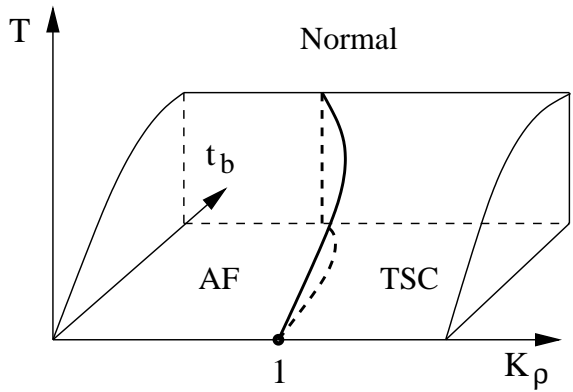}
\caption{Proposed phase diagram for weak inter-chain coupling
$t_b$.  When inter-chain coupling is present, $t_b \ne 0$, long
range order at finite temperatures becomes possible.  In the
unitary case, the second order quantum critical point of a
Luttinger liquid becomes a first order transition between AF and
unitary TSC.  As $t_b$ grows, the AF phase shrinks due to reduced
nesting of the Fermi surface. Throughout we assume positive
backscattering $g_1>0$. \label{quasi1d}}
\end{figure}

We illustrate the effects of interchain coupling in Fig.
\ref{quasi1d}.  As the temperature is reduced on either side of
$K_\rho=1$, the correlation length along a chain in the
appropriate correlation function grows due to intrachain
interactions.  At some finite temperature, before this length
diverges, coupling between the chains becomes relevant and a true
three dimensional transition can take place. At the crossover
between one and three dimensions, the description of the system in
terms of a Luttinger liquid is supplanted by a GL free energy
describing the interactions of the order parameters in three
spatial dimensions. In this picture, for small enough $t_b$ the
presence of a phase boundary between AF and TSC implies that the
intrachain Hamiltonian is close to the SO(4) symmetric point
$K_\rho=1$.

An important assumption of our analysis is that pressure varies
the value of $K_\rho$ and tunes the transition across the AF/TSC
phase boundary.  We note that measurements of $K_\rho$ based on
optical conductivity measurements have been carried out at ambient
pressure, {\it i.e.} deep inside the SDW phase in
(TMTSF)$_2$PF$_6$. For instance, in Refs.~\onlinecite{Degiorgi}
and \onlinecite{Schwartz1998}, the value $K_\rho= 0.23$ is
obtained. Ref.~\onlinecite{Schwartz1998} point out that this value
of $K_\rho$ assumes that the dominant umklapp contribution is due
to commensurability at quarter filling. They acknowledge that if
umklapp is dominated by commensurability at half-filling, these
measurements then imply $K_\rho=0.925$. They also point out that a
half-filled model has the drawback of predicting a gap energy that
is too small. However, it is equally difficult to justify the
assumption that quarter filling commensurability is dominant: for
$K_\rho=0.23$, the rate of divergence in the RG of
commensurability at one quarter is exponentially smaller than the
corresponding rate for commensurability at half-filling.  Thus, we
feel that the question of the value of $K_\rho$ is not yet
settled.

\subsection{Incommensurate filling}
\label{incommensurateSection}

At incommensurate filling, a translation by one lattice constant
multiplies the SDW order parameter by a complex phase factor, \be
\vec{\Phi}\to e^{2i{\bf k_f}\cdot {\bf a}}\vec{\Phi}= e^{2\pi
i\nu}\vec{\Phi}, \ee where $\nu$ is the filling fraction of the
conduction band.  For a completely incommensurate case, when $\nu$ is
an irrational number, the GL free energy must be SO(2) symmetric
with respect to the phase of $\Phi$\cite{Zhang2002,endnotePin}. In the
absence of pinning terms, the most general GL free energy with
SO(3)$_{\rm spin}\times$SO(2)$_{\rm charge}\times$SO(2)$_{\rm
translation}$ is
\begin{eqnarray}
F&=&
\frac{1}{2}\,|\, \nabla \vec{\Psi}\,|^2 + \frac{1}{2}\, |\nabla \vec{\Phi}|^2
+ \frac{r_1}{2} |\vec{\Psi}|^2 + \frac{r_2}{2} |\vec{\Phi}|^2 +u_1
(|\vec{\Psi}|^2)^2 + u_2 (|\vec{\Phi}|^2)^2 \nn\\
&+& u_3 |\vec{\Psi}^2|^2
+u_4 |\vec{\Phi}^2|^2+ 2 v_1 |\vec{\Psi}|^2 |\vec{\Phi}|^2
+ 2 v_2 | \vec{\Phi}\cdot \vec{\Psi}|^2+2 v_3 | \vec{\Phi}^*\cdot \vec{\Psi}|^2
\label{incomGLenergy}
\end{eqnarray}
Near the phase boundary between SDW and TSC phases, for quasi one
dimensional systems the form of the free energy is strongly
constrained by the SO(3)$_{\rm spin}\times$SO(4)$_{\rm isospin}$
symmetry. We expect the properties of the system to be well
described by the free energy,
\begin{eqnarray}
F &=& \frac{1}{2} \sum_{\bar{a}\alpha} \nabla P_{\bar{a}\alpha}
\nabla P_{\bar{a}\alpha}+
\frac{\bar{r}}{2}\sum_{\bar{a}\alpha} P_{\bar{a}\alpha} P_{\bar{a}\alpha}
+\frac{\delta r}{2}\sum_\alpha \left(P_{1\a}^2+P_{2\a}^2-P_{3\a}^2-P_{4\a}^2\right)\nn\\
&+&\tilde{u}_1 \sum_{\bar{a} \alpha \bar{b} \beta } P_{\bar{a}\alpha} P_{\bar{a}\alpha}
P_{\bar{b} \beta} P_{\bar{b} \beta}
+\tilde{u}_2 \sum_{\bar{a} \alpha \bar{b} \beta } P_{\bar{a}\alpha} P_{\bar{a}\beta}
P_{\bar{b} \alpha} P_{\bar{b} \beta}.
\label{GinvGL}
\end{eqnarray}
This is the most general free energy with
SO(3)$_{\rm spin}\times$SO(4)$_{\rm isospin}$ symmetric quartic coefficients,
where we have used the order parameter defined in equation (\ref{Pmatrix}) to
display this invariance explicitly.  We follow the common assumption that
changing the external control parameters of the system
only affects the quadratic coefficients.  Thus, these are allowed to break
the symmetry and tune the phase transition.  For $\delta r\ne 0$,
the symmetry is broken down to
SO(3)$_{\rm spin}\times$SO(2)$_{\rm charge}\times$SO(2)$_{\rm translation}$.

\begin{figure}
\includegraphics[width=5cm]{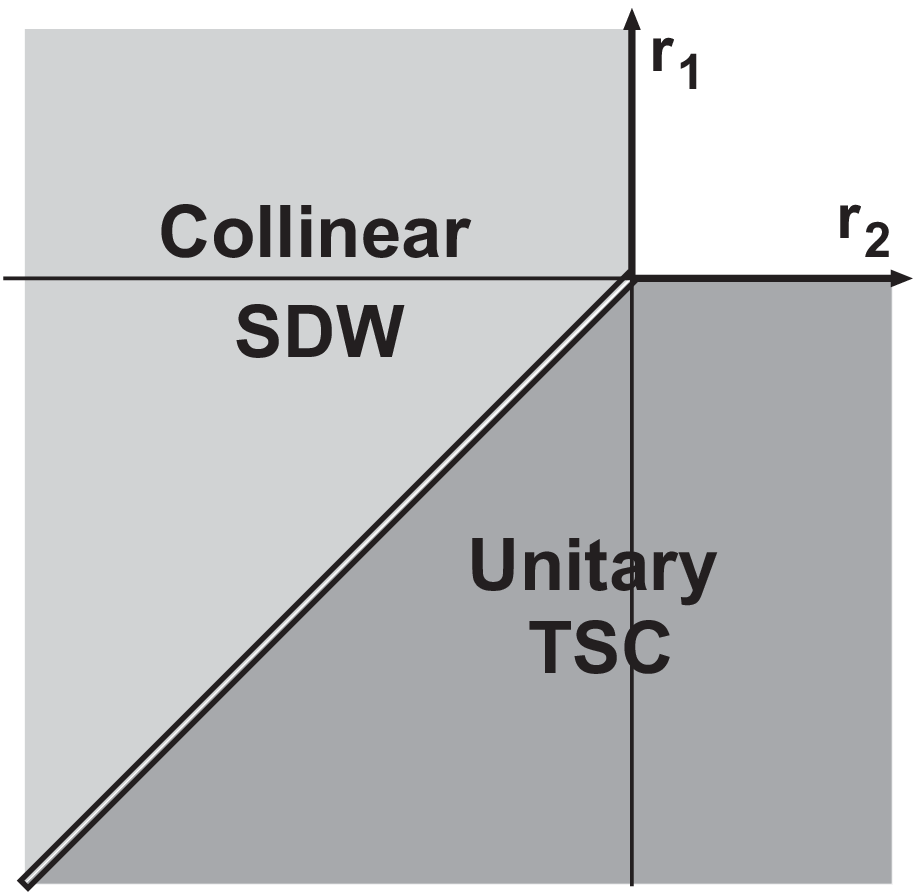}
\caption{Mean-field phase diagram of (\ref{GinvGL}) and (\ref{NearlySO4GLenergy}) for
$\tilde{u}_2<0$.  At half filling, SDW order reduces to AF order.
\label{figNearlySO4U2Negative}}
\end{figure}
\begin{figure}
\includegraphics[width=5cm]{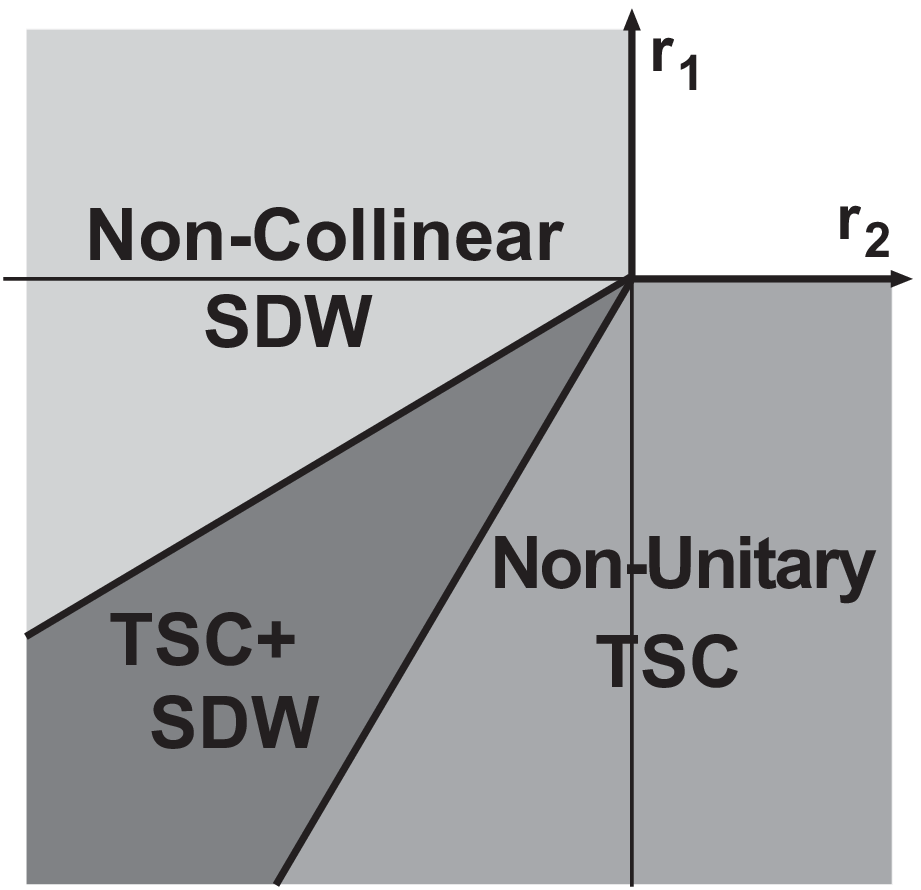}
\caption{Mean-field phase diagram of (\ref{GinvGL}) and (\ref{NearlySO4GLenergy})
for $\tilde{u}_2>0$. At half filling, SDW order reduces to AF order.
\label{figNearlySO4U2Positive}}
\end{figure}

There is an explicit duality between antiferromagnetism and
triplet superconductivity under reversal of the sign of $\delta r$
in (\ref{GinvGL}).  The mean field phase diagram of (\ref{GinvGL})
depends crucially on the sign of $\tilde{u}_2$.  For negative
$\tilde{u}_2$, there is a tendency for all vector order parameters
to point along a common axis.   The order parameters in this case
can be described by a single real vector times a complex phase,
$\vec{\Psi}=e^{i\varphi}\vec{n}$ and
$\vec{\Phi}=e^{i\theta}\vec{n}$. This is referred to in the $^3He$
literature as {\it unitary} triplet
superconductivity\cite{LeggettRMP}, and in magnetism as {\it
collinear} spin density order\cite{Zhang2002}. On the other hand,
for positive $\tilde{u}_2$, all vector order parameters tend to be
mutually orthogonal.  The real and imaginary parts of the order
parameters can no longer be set to be parallel,
Re$\vec{\Psi}\times$Im$\vec{\Psi}\ne 0$ and
Re$\vec{\Phi}\times$Im$\vec{\Phi}\ne 0$. This is the {\it
non-unitary}/{\it non-collinear} case. The mean-field phase
diagrams for (\ref{GinvGL}) for $\tilde{u}_2$ negative and
positive are shown in Figs. \ref{figNearlySO4U2Negative} and
\ref{figNearlySO4U2Positive}.

In principle, the parameters of the GL free energy
(\ref{incomGLenergy}) can be obtained from a microscopic
Hamiltonian. In Appendix \ref{AppendixGLDerivation} we consider a
quasi one dimensional electron systems with weak interactions and
obtain a free energy as in (\ref{GinvGL}) with
\begin{eqnarray}
\tilde{u}_1 &=& \frac{21 \zeta (3)}{16\pi^2 v_f T^2}
\nonumber\\
\tilde{u}_2 &=& -\frac{7 \zeta (3)}{8\pi^2 v_f T^2}
\label{MFtildeu}
\end{eqnarray}
The quadratic coefficients depend on coupling constants in the TSC
and SDW channels, and on temperature. They are typically parameterized in
terms of the pressure-dependent mean-field transition temperature
\begin{eqnarray}
r_1(T,P) &=& \alpha_{\rm TSC}(T-T_C(P))
\nonumber\\
r_2(T,P) &=& \alpha_{\rm SDW}(T-T_N(P))
\end{eqnarray}
The analysis of Appendix \ref{AppendixGLDerivation} shows that weakly interacting Fermi
liquids favor unitary TSC and collinear SDW order, $\tilde{u}_2<0$.

\subsection{Half-filling}
\label{SectionHalfFillingGL}

As is well known, the physics of period two antiferromagnetic
order is captured by a single real N\'eel vector.  It is
interesting to study how this comes about from the point of view
of the microscopic Luttinger Hamiltonian.  As pointed out in
Section \ref{LuttingerSection}, the Hamiltonian at half-filling
includes a new contribution due to umklapp scattering, Eq.
(\ref{umkSDW}). This gives a correction to the GL free energy,
which to linear order in $g_3$ can be written as
\begin{eqnarray}
\Delta F=h\left(({\rm Re}\vec{\Phi})^2-({\rm
Im}\vec{\Phi})^2\right) \label{pinning}
\end{eqnarray}
where $h=\frac{g_3}{2L}$.  The new term pins
the SDW, and breaks the SO(4)$_{\rm isospin}$=SO(3)$_R
\times$SO(3)$_L$ symmetry down to its diagonal subgroup
SO(3)$_{\rm isospin}$.  In agreement with the Feynman-Hellman
theorem, the quartic coefficients derived in Appendix
\ref{AppendixGLDerivation} are not modified to linear order in
$g_3$. Therefore, in the linear order in $g_3$, the free energy
has SO(4)=SO(3)$_{\rm spin}\times$SO(3)$_{\rm isospin}$ symmetry.
In principle, the higher order contributions of the umklapp $g_3$
term can break the original SO(4)$_{\rm isospin}$ symmetry all the
way down to SO(2)$_{\rm charge}$, generated by the total charge
$Q$.

The bare value of $g_3$ in (TMTSF)$_2$PF$_6$ is small, since it is
proportional to dimerization, which in this compound is very
weak\cite{Thorup}.   Assuming that Coulomb interactions
are of the order of the bandwidth, this leads to a bare value of $g_3$
of about $0.01$. Furthermore, it is not the bare value of $g_3$
that enters the GL free energy, but its effective (renormalized)
value at the 1d to 3d crossover scale. At high temperatures
one-dimensional physics is observed.  As temperature is reduced,
everywhere on the TSC side of the phase diagram, as well as on the AF/TSC
phase boundary, $g_3$ flows to zero.  This allows us to
approach the critical region from the TSC side, along which the
GL free energy is SO(3)$_{\rm iso}$ symmetric.  Even on the AF side of the
phase diagram, where $g_3$ is relevant, the flow of $g_3$ passes
near zero before diverging.  Therefore, near the AF/TSC phase boundary, the
flow spends a lot of time near zero, and the eventual upturn of
$g_3$ may not be reached for realistic systems, in which the 3d
coupling may cut off the 1d RG flow at low temperatures. Hence, it
is reasonable to take small $g_3$ everywhere near the AF/TSC phase
boundary, and to consider a model with SO(3)$_{\rm iso}$ symmetry.

In what follows, we will assume that the umklapp term favors the
real part of the SDW, which becomes the N\'eel order parameter
$\vec{N}$,
\be
\vec{N}={\rm Re}\vec{\Phi}.
\ee
From now on we assume that Im$\vec{\Phi}$ is sufficiently well
gapped, so that it does not need to be included in the analysis of
the competition between AF and TSC.  This is justified since the
pinning term (\ref{pinning}) is relevant in the 3d theory, and
thus any fixed point of the theory will be characterized by strong
pinning.

It is useful to consider the relation between our SO(4) symmetry
and the SO(4) symmetry introduced by C.N. Yang for the Hubbard
model\cite{Yang1990}.  The symmetry generators of Yang's SO(4) is
the $\eta$ operator,
\begin{eqnarray}
\eta^\dagger=\sum_k
c^\dagger_{k+\pi\uparrow}c^\dagger_{-k\downarrow}.
\end{eqnarray}
$k$ summation goes over the entire Brillouin zone. This operator
should be compared to our $\Theta$ operators defined in eq.
(\ref{ThetaOperator}) for incommensurate filling, and eq.
(\ref{quantumGenerators}) for half-filling.  Away from half
filling, the difference between the two operators is obvious.
$\Theta$ has momentum equal to the nesting wave vector $2k_f$, in
contrast to $\eta$, which always has center of mass momentum
$\pi$.  This allows us to have SO(4) symmetry for any electron
density, in contrast to Yang's SO(4), which only applies at
half-filling.  At half-filling, however, $2k_f=\pi$, and the only
difference is the relative sign between the left and right moving
contributions. This difference is substantial.  The N\'eel order
parameter transforms as a {\it singlet} under the action of
$\eta$,
\begin{eqnarray}
\left[ \eta, \vec{N} \right]=0.
\end{eqnarray}
This should be contrasted with $\Theta$, which rotates $\vec{N}$
into the TSC order parameter $\vec{\Psi}$, see eq.
(\ref{OrderParameterLuttinger}).  The two SO(4) symmetries thus
differ in the order parameters which they unify, and in the
microscopic models for which they apply.
Yang's SO(4) applies to the negative $U$ Hubbard model, for which
singlet SC and CDW are degenerate lowest energy states at
half-filling. Our SO(4) unifies AF and TSC orders, which are not
degenerate for the Hubbard model.  As we discussed earlier, we
expect these to be nearly degenerate phases for half-filled
systems with small umklapp (e.g. quarter filled systems with small
dimerization, such as (TMTSF$_2$)PF$_6$ and with $K_\rho$ close to
one).


In analogy with the incommensurate case, at half-filling we expect that quasi
one-dimensional systems near the AF/TSC phase
boundary have a Ginzburg-Landau free energy with SO(4)-symmetric quartic coefficients.  The
symmetry can be made explicit in terms of the matrix order parameter $\hat{Q}$:
\begin{eqnarray}
F &=& \frac{1}{2} \sum_{a\alpha} \nabla Q_{a\alpha} \nabla
Q_{a\alpha}+ \frac{\hat{r}}{2}\sum_{a\alpha} Q_{a\alpha}
Q_{a\alpha} +\delta r \sum_\a(Q_{3\a}^2-Q_{1\a}^2-Q_{2\a}^2)\nn\\
&\,&+\tilde{u}_1 \sum_{a \alpha b \beta } Q_{a\alpha} Q_{a\alpha}
Q_{b \beta} Q_{b \beta} +\tilde{u}_2 \sum_{a \alpha b \beta }
Q_{a\alpha} Q_{a\beta} Q_{b \alpha} Q_{b \beta}
\nonumber\\
&=&
\frac{1}{2}\,|\, \nabla \vec{\Psi}\,|^2 + \frac{1}{2}\, (\nabla \vec{N})^2
+
\frac{r_1}{2} |\vec{\Psi}|^2 + \frac{r_2}{2} \vec{N}^2
\nonumber\\
&+&
(\tilde{u}_1+\frac{\tilde{u}_2}{2})(|\vec{\Psi}|^2)^2
+\frac{\tilde{u}_2}{2} |\vec{\Psi}^2|^2
+2 \tilde{u}_1  |\vec{\Psi}|^2 \vec{N}^2
+2 \tilde{u}_2 |\vec{\Psi}\cdot\vec{N}|^2
+(\tilde{u}_1+\tilde{u}_2) (\vec{N}^2)^2
\label{NearlySO4GLenergy}
\end{eqnarray}
Changing temperature and some other parameter of the system
(e.g. pressure in (TMTSF)$_2$PF$_6$) allows to control $r_1$ and $r_2$.
The SO(4) symmetry is recovered on the line $r_1=r_2$.

Equation (\ref{NearlySO4GLenergy}) is a special case of
the most general free energy with the $SO(3)\times SO(2)$
symmetry of spin and charge rotations\cite{endnoteTSC},
\begin{eqnarray}
\tilde{F}=
\frac{1}{2}\,|\, \nabla \vec{\Psi}\,|^2 + \frac{1}{2}\, (\nabla \vec{N})^2
+ \frac{r_1}{2} |\vec{\Psi}|^2 + \frac{r_2}{2} \vec{N}^2 +u_1
(|\vec{\Psi}|^2)^2 + u_2 (\vec{N}^2)^2 + u_3 |\vec{\Psi}^2|^2
+ 2 v_1 |\vec{\Psi}|^2 \vec{N}^2
+ 2 v_2 | \vec{N}\cdot \vec{\Psi}|^2
\label{GLenergy}
\end{eqnarray}
Translational symmetry rules out $\vec{N} \cdot \vec{\Psi}^* \times
\vec{\Psi} $ because this term has a non-zero wave vector.  Similarly,
$| \vec{N} \times \vec{\Psi}|^2$ can be reduced
to terms already present in (\ref{GLenergy}).
When the conditions
\begin{eqnarray}
r_1 &=& r_2  \nonumber\\
u_2-u_3&=&u_1 \nonumber\\
u_2-2u_3&=&v_1 \nn\\
v_2&=&2u_3\label{SO4Condition}
\end{eqnarray}
are satisfied, we recover SO(4) symmetry.    In addition, if we
supplement the conditions (\ref{SO4Condition}) by
\begin{eqnarray}
v_2=0 \label{SO9condition}
\end{eqnarray}
there is an even higher SO(9) symmetry, which allows rotations between
any components of vectors $ \vec{N}$, ${\rm Re} \vec{\Psi}$, and
${\rm Im} \vec{\Psi}$
\begin{eqnarray}
\bar{F} &=&
\frac{1}{2}\,|\, \nabla \vec{\Psi}\,|^2
+ \frac{1}{2}\, (\nabla \vec{N})^2
+\frac{r}{2} ( |\vec{\Psi}|^2 + \vec{N}^2 )
+ \bar{u} (|\vec{\Psi}|^2 + \vec{N}^2 )^2
\label{SO9GLenergy}
\end{eqnarray}

At half-filling, there is no distinction between collinear and non-collinear
magnetism.  However, the sign of $\tilde{u}_2$ still determines the
nature of the triplet superconductivity, as well as the topology of the
mean-field phase diagrams.  These are similar to those displayed for
incommensurate filling, Figs. \ref{figNearlySO4U2Negative}
and \ref{figNearlySO4U2Positive}, the only difference being
that SDW order is reduced to AF order.
The mean-field phase diagrams for
(\ref{NearlySO4GLenergy}) for $\tilde{u}_2$ negative and
positive are shown in Fig. \ref{figNearlySO4U2Negative}.
The analysis of Appendix \ref{AppendixGLDerivation} can be easily
modified to half-filling, and yields an SO(4) symmetric free energy of the form
(\ref{NearlySO4GLenergy}) with coefficients still given by (\ref{MFtildeu}).
Thus, weekly interacting Fermi liquids favor the case $\tilde{u}_2<0$.
Strong interactions, however, can modify
the quartic coefficients (\ref{NearlySO4GLenergy}), including a possible
change of sign of $\tilde{u}_2$. In the subsequent discussion we
consider both possibilities. It is useful to note that all
experimentally known cases of triplet pairing between fermions, such
as $^{3}He$ \cite{LeggettRMP,PWA1984}, and $Sr_2RuO_4$, correspond to
the unitary case. Hence, negative $\tilde{u}_2$ appears more likely.

\section{Thermal fluctuations}
\label{SectionThermal}

We now consider the free energy (\ref{GLenergy}) and address how
fluctuations affect the mean-field phase diagram shown in Figs.
\ref{figNearlySO4U2Negative} and \ref{figNearlySO4U2Positive}. For
instance, when the quartic coefficients do not lie exactly on the
SO(4) symmetric manifold, we will study whether such symmetry
appears as we go to longer length scales and integrate out short
wave length fluctuations. The possibility of enhanced static
symmetry at the critical point has been discussed previously for
several solid state systems. For easy axis AF in a magnetic field,
SO(3) symmetry was suggested to appear at the spin flop critical
point \cite{Fisher1974,Kosterlitz1976}. For systems with competing
singlet superconducting and antiferromagnetic orders Zhang
suggested a static SO(5) symmetry as the bicritical
point\cite{Zhang1997,DemlerRMP}.  This SO(5) symmetry has also
been used to study the quasi-two dimensional $\kappa$-BEDT-TTF
salts\cite{Murakami2000}.  Yang and Zhang introduced an SO(4)
symmetry for the Hubbard model at half-filling which unifies
singlet superconductivity with charge density wave order
\cite{Yang1989,Zhang1990}.

To understand the role of fluctuations in models (\ref{GLenergy})
and (\ref{NearlySO4GLenergy}) we use $4-\epsilon$ renormalization
group (RG) analysis.  For subsequent discussion it is useful to
extend the spin SO(3) symmetry of the equation (\ref{GLenergy}) to
a more general SO(N) symmetry. This is achieved by considering
vectors $\vec{N}$ and $\vec{\Psi}$ as $N$-component vectors.  The
RG equations can be derived using the standard approach
\cite{Chaikin}
\begin{eqnarray}
\frac{d r_1}{dl} &=& 2r_1+\frac{8K_d}{1+r_1}\{
(N+1)u_1+3u_3\}+\frac{4K_d}{1+r_2}\{Nv_1+2v_2\}
 \nonumber\\
\frac{d r_2}{dl} &=& 2r_2+\frac{8K_d}{1+r_1}\{
Nv_1+2v_2\}+\frac{4K_d}{1+r_2}(N+2)u_2 \nonumber\\
\frac{d u_1}{dl} &=& \epsilon u_1 - K_d \{
(8N+32)u_1^2+32u_1u_3+32u_3^2+4Nv_1^2+8v_1v_2+2v_2^2 \}
\nonumber\\
\frac{d u_2}{dl} &=& \epsilon u_2 - K_d \{
4(N+8)u_2^2+8Nv_1^2+16v_1v_2+8v_2^2
\}\nonumber\\
\frac{d u_3}{dl} &=& \epsilon u_3 - K_d \{ 8Nu_3^2+48u_1u_3+2v_2^2
\}\nonumber\\
\frac{d v_1}{dl} &=& \epsilon v_1 - K_d \{ (8N+8)u_1v_1 +(4N+8)u_2
v_1+16u_3v_1 +16v_1^2+8u_1 v_2+4v_2u_2+4v_2^2
\}\nonumber\\
\frac{d v_2}{dl} &=& \epsilon v_2 - K_d \{ 8u_1
v_2+8u_2v_2+16u_3v_2+32v_1v_2+(4N+8)v_2^2 \}
\label{4-eRGequations}
\end{eqnarray}
Here $dl=d\Lambda/\Lambda$, where $\Lambda$ is a momentum cut-off,
and $K_d=\dots$ is a surface of a unit sphere in $d=4-\epsilon$
dimension.

\begin{figure}
\includegraphics[width=7.5cm]{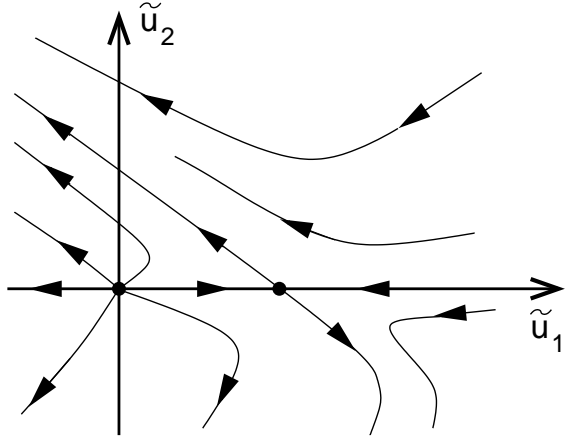}
\caption{Renormalization group flow of the SO(4) symmetric theory
eq. (\ref{NearlySO4GLenergy}) in $d=4-\epsilon$ dimensions.  The
sign of $\tilde{u}_2$ does not change under the flow, and there
are no stable fixed points.  Instead, there are two types of
runaway flow, corresponding to unitary ($\tilde{u}_2<0$) and
non-unitary ($\tilde{u}_2>0$) TSC.  The two are separated by a
line of $SO(9)$ symmetric theories ($\tilde{u}_2=0$).
\label{symflow}}
\end{figure}

For the physically relevant $N=3$ equations (\ref{4-eRGequations})
have only two fixed points. One is a trivial Gaussian fixed point
\begin{eqnarray}
r_{\{1,2\}}=u_{\{1,2,3\}}=v_{\{1,2\}}=0
\end{eqnarray}
and the other is an SO(9) Heisenberg point
\begin{eqnarray}
r_{\{1,2\}}&=&-\frac{(3N+2)\epsilon}{6N+16}
\nonumber\\
u_1=u_2=v_1&=&\frac{\epsilon}{(6N+16)K_d}
\nonumber\\
u_3=v_2&=&0
\end{eqnarray}
The Gaussian fixed point is completely unstable. The SO(9) Heisenberg
point has five unstable directions (for general $N$,
the Heisenberg point has $SO(3N)$ symmetry, but it remains unstable
in five directions for all $N>1$).
The critical point should have only two unstable directions:
$r_{\{1,2\}}$ should flow away from the critical point, but all
the interaction coefficient should flow toward the fixed point.
So, neither the Gaussian nor the SO(9) Heisenberg fixed points are
good candidates for the critical point. In Fig. \ref{symflow} we
show RG flows in the SO(4) symmetric plane. We find two types of
run-away flows.  When we start with $\tilde{u}_2$ positive, it
continues increasing. For $\tilde{u}_2$ negative, the RG flow make
it even more negative. In both cases $\tilde{u}_1$ flows to negative
values.

In many cases absence of a fixed point in the RG flows implies that we
do not have a multicritical point in the phase diagram, but instead
fluctuations induce a first order phase transition. Below we discuss
consequences of the run-away flows in equations
(\ref{4-eRGequations}).  We point out that two types of the run-away
flows in the SO(4) symmetric manifold shown in Fig. \ref{symflow}
correspond to unitary ($\tilde{u}_2<0$) and non-unitary
($\tilde{u}_2>0$) TSC. These two cases are considered
separately.

\section{Finite temperature analysis:
Unitary case
}
\label{SectionUnitaryTSC}

We consider model (\ref{NearlySO4GLenergy}) with $N$-component vectors
and with negative $\tilde{u}_2$. In the $4-\epsilon$ expansion there are
run-away flows even for large $N$. Thus, we do not find a fixed point
that could give a critical point.  To understand the phase diagram
in this case we employ large $N$ calculations in $d=3$.  In the large $N$ expansion,
all bubble diagrams are summed self-consistently\cite{ZinnJustin,Ma}.
The large $N$ approach for unitary triplet
superconductors without coupling to magnetic order has been discussed
previously in \cite{Bailin1977} in the context of $^{3}{\rm He}$.

Let us start by analyzing the superconducting phase. In the mean-field
approximation the order parameter factorizes as
$\vec{\Psi}=e^{i\theta}\vec{n}$. Hence, we take the average value of
the order parameter in the ordered phase to be $\vec{\Psi}_0 =
(0,...,0,\sigma)$ and separate the longitudinal and transverse components
of the fluctuating part  $\delta \vec{\Psi}=(\vec{A}_T+i\vec{B}_T,
A_L+iB_L)$. For the N\'eel order parameter we also separate
$\vec{N}=(\vec{N}_T,N_L)$. Effective masses for $\vec{A}_T$,
$\vec{B}_T$, and $\vec{N}_T$ are given by
\begin{eqnarray}
r_A&=&r_1+4(\tilde{u}_1 + \tilde{u}_2) \sigma^2 + 4(\tilde{u}_1 +
\tilde{u}_2) N \int_0^\Lambda \frac{d^3k}{(2\pi)^3}\,
\frac{1}{k^2+r_A}+ 4\tilde{u}_1  N  \int_0^\Lambda
\frac{d^3k}{(2\pi)^3}\, \left( \frac{1}{k^2+r_B} +
\frac{1}{k^2+r_N} \right)
\nonumber\\
r_B&=&r_1 + 4\tilde{u}_1 \sigma^2
+ 4(\tilde{u}_1 + \tilde{u}_2) N  \int_0^\Lambda \frac{d^3k}{(2\pi)^3}\,
\frac{1}{k^2+r_B}
+ 4\tilde{u}_1  N \int_0^\Lambda \frac{d^3k}{(2\pi)^3}\,
\left( \frac{1}{k^2+r_A} + \frac{1}{k^2+r_N} \right)
\nonumber\\
r_N&=&r_2 + 4\tilde{u}_1 \sigma^2
+ 4(\tilde{u}_1 + \tilde{u}_2)  N \int_0^\Lambda \frac{d^3k}{(2\pi)^3}\,
\frac{1}{k^2+r_N}
+ 4\tilde{u}_1  N \int_0^\Lambda \frac{d^3k}{(2\pi)^3}\,
\left( \frac{1}{k^2+r_A} +  \frac{1}{k^2+r_B} \right)
\label{TransverseMasses}
\end{eqnarray}
where $\Lambda$ is the ultraviolet (short distance)
cut-off of the free energy in equation (\ref{NearlySO4GLenergy}).
In writing equations (\ref{TransverseMasses}) we used that in the
large $N$ limit, $\tilde{u}_{\{1,2\}} \sim 1/N$, $\sigma \sim
\sqrt{N}$, and we neglected terms of the order of $1/N$, including
contributions from longitudinal fluctuations.  A requirement of the
cancellation of tadpole diagrams for $A_L$ gives the condition
$r_A=0$, as one would expect from the Goldstone theorem.  It is
convenient to define parameter $r_c$ from the condition
\begin{eqnarray}
0=r_c+(12 \tilde{u}_1 + 4 \tilde{u}_2)N \int_0^\Lambda \frac{d^3k}{(2\pi)^3}\,
\frac{1}{k^2}
\label{rcdefinition}
\end{eqnarray}
If we measure $r$'s with respect to $r_c$
\begin{eqnarray}
t_\psi &=& r_1-r_c
\nonumber\\
t_N &=& r_2-r_c
\label{EqnshiftedRs}
\end{eqnarray}
we can absorb all the cut-off dependence of equations
(\ref{TransverseMasses}) into definitions of $t_\psi $ and $t_N$
\begin{eqnarray}
0 &=& t_\psi +4(\tilde{u}_1 + \tilde{u}_2) \sigma^2
+ 4 \tilde{u}_1 N \int \frac{d^3k}{(2\pi)^3}\,
\left( \frac{1}{k^2} - \frac{1}{k^2+r_B} \right)
\nonumber\\
r_B &=& t_\psi +4\tilde{u}_1 \sigma^2
+4(\tilde{u}_1 + \tilde{u}_2)  N\int
\left( \frac{1}{k^2} - \frac{1}{k^2+r_B} \right)
+ 4 \tilde{u}_1 N \int \frac{d^3k}{(2\pi)^3}\,
\left( \frac{1}{k^2} - \frac{1}{k^2+r_N} \right)
\nonumber\\
r_N &=& t_N +4\tilde{u}_1 \sigma^2
+  4 \tilde{u}_1 N \int \frac{d^3k}{(2\pi)^3}\,
\left( \frac{1}{k^2} - \frac{1}{k^2+r_B} \right)
+4(\tilde{u}_1 + \tilde{u}_2)  N \int
\left( \frac{1}{k^2} - \frac{1}{k^2+r_N} \right)
\label{NTSCequations}
\end{eqnarray}
Integrals are now convergent for large $k$, so
upper limits of integration can be extended to infinity.
Solutions to equations (\ref{NTSCequations}) correspond to extremal
points of free energy as a function of $\sigma$.
When both $t_\psi$ and $t_N$ are large, there are no solutions to
equations (\ref{NTSCequations}). This is a disordered phase. Once we
decrease $t_\psi$ sufficiently, a single solution appears at
$t_{\psi,M}$ and splits into two for $t_\psi<t_{\psi,M}$. This
describes the appearance of the TSC phase as a locally stable
state. The point $t_{\psi,M}$, where two solutions merge into one and
disappear, correspond to the boundary of the  local stability region
of the TSC phase. As $t_\psi$ is lowered further, at a temperature
$t_{\psi,L}$ one of the solutions approaches $\sigma=0$ and then
disappears. This is a spinodal point below which a disordered phase is
unstable to developing TSC order parameter. The actual first order
phase transition occurs somewhere between $t_{\psi,M}$ and
$t_{\psi,L}$.

\begin{figure}
\includegraphics[width=14cm]{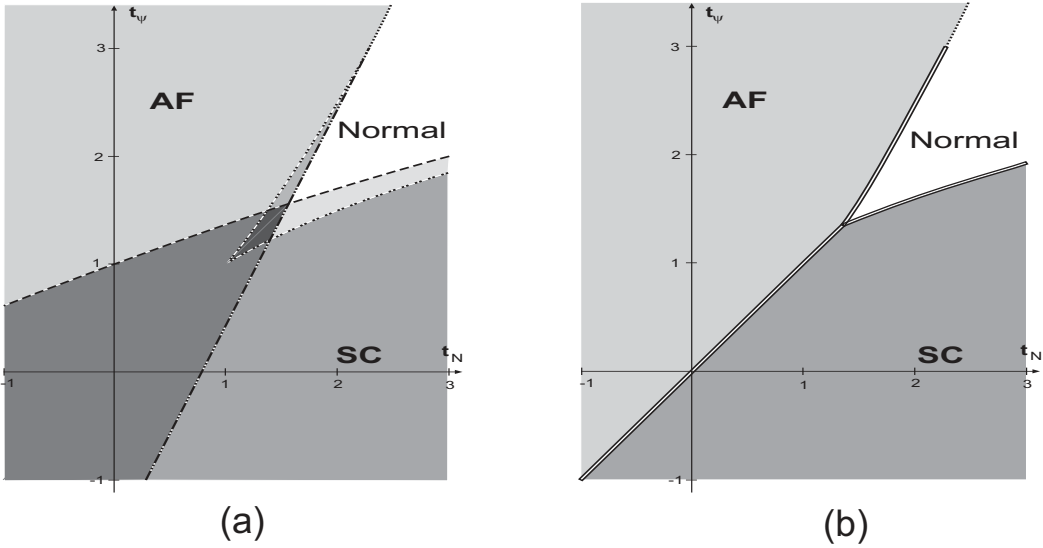}
\caption{Phase diagram of (\ref{NearlySO4GLenergy})
with $\tilde{u}_2<0$ in three dimensions in the large
$N$ limit including fluctuations. Parameters are $\tilde{u}_1=1/N$
$\tilde{u}_2=-1/2N$.
\label{figLargeNU2Negative}}
\end{figure}

Fig. \ref{figLargeNU2Negative} shows a phase diagram
constructed from the arguments presented
above, for both TSC and AF phases. We note that the mixed phase with
simultaneous TSC and AF orders is only possible on the $t_\psi=t_N$
line. Thus, the first order phase transition between two types of
ordered phases remains even when we include fluctuations.
The most interesting feature of this phase diagram is that the
transition between the disordered and the antiferromagnetic phases
becomes first order in the vicinity of the critical point.

\section{Finite temperature analysis: non-unitary case
}
\label{SectionNonUnitary}

Mean-field calculations for the free energy (\ref{NearlySO4GLenergy})
with $\tilde{u}_2>0$ demonstrated that the TSC phase is non-unitary
and there is a mixed phase with both TSC and AF order
(see Section \ref{SectionGLDiscussion} and
figure \ref{figNearlySO4U2Positive}). Within mean field theory,
the mixed phase terminates at a tetracritical point with $SO(4)=SO(3)\times SO(3)$
symmetry. The goal of this section is to examine
how the tetracritical point is affected by thermal fluctuations.

To this end we extend the SO(3) spin symmetry to SO($N$) and
approach the problem with three different methods: A large
$N$ analysis, a renormalization group calculation in $d=4-\e$ and in one in $d=2+\e$.
The physical picture that emerges form all of these approaches
is that a $SO(3)\times SO(N)$ critical point exists for sufficiently large $N$,
but probably does not survive down to the physical $N=3$. We argue that in this
case the tetracritical point is stretched to a line of direct first order transition
from the normal state to the mixed phase.

\subsection{Large $N$ phase diagram in three dimensions}
\label{SectionNonUnitaryFiniteTLargeN}

We consider the model (\ref{NearlySO4GLenergy})
in three dimensions for large $N$ and with $\tilde{u}_2>0$.
We note that the quartic terms give a free energy
that is bounded from below for $\tilde{u}_1+\tilde{u}_2/3>0$.
Thus, in this section we will always assume that
this condition is satisfied. In appendix \ref{AppendixB}
we also discuss that when $\tilde{u}_1+\tilde{u}_2/3$
becomes small, of the order of $1/N^2$ (in the
large $N$ limit both $\tilde{u}$s are of the order
of $1/N$), the phase diagram may change qualitatively.

\begin{figure}
\includegraphics[width=6cm]{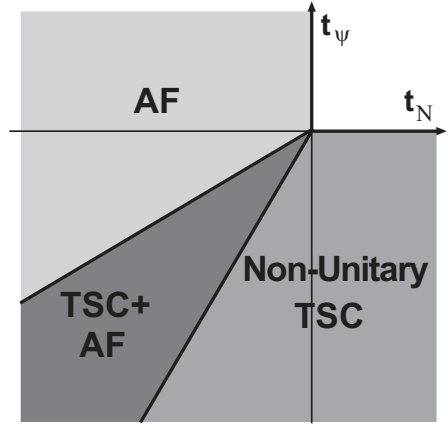}
\caption{Phase diagram of the model  (\ref{NearlySO4GLenergy})
with $\tilde{u}_2>0$
in three dimensions in the large $N$ limit. The four second order
lines meet at the tetracritical point at non-zero angles.
The same phase diagram appears in the
$d=4-\epsilon$ analysis for $N\ge 33$ and
in the $d=2+\epsilon$ analysis for $N \ge 5$.
\label{figNearlySO4U2PositiveFluctuations}}
\end{figure}

 In the mixed phase
the TSC and AF both have non-zero expectation values and are
orthogonal to each other. Hence, in the ordered phase
we can choose
\begin{eqnarray}
\langle \vec{\Psi} \rangle &=& (0,\dots,0,\sigma_\psi,i \sigma_\psi,0)
\nonumber\\
\langle \vec{N} \rangle &=& (0,\dots,0,0,0,\sigma_N)
\end{eqnarray}
Following the discussion in Section \ref{SectionUnitaryTSC} we
introduce longitudinal and transverse fluctuations for all order
parameters. It is easy to verify that the requirement of
cancellation of tadpole diagrams for longitudinal components
implies zero effective masses for the transverse components.
Shifting $r_{1}$ and $r_{2}$ as in equation (\ref{EqnshiftedRs})
we obtain self-consistency conditions for expectation
values of the order parameters
\begin{eqnarray}
t_\psi+(12\tilde{u}_1+4\tilde{u}_2) \sigma_\psi^2
+4 \tilde{u}_1 \sigma_N^2 &=&0
\nonumber\\
t_\psi+(4\tilde{u}_1+4\tilde{u}_2) \sigma_N^2
+8 \tilde{u}_1 \sigma_\psi^2 &=&0
\end{eqnarray}
These equations can be easily solved and we obtain a phase diagram
shown in figure \ref{figNearlySO4U2PositiveFluctuations}.
We observe that in this case the only effect
of fluctuations is to shift the tetracritical point from $r_1=r_2=0$
to $r_1=r_2=r_c$.

\subsection{Renormalization Group Analysis in $d=4-\epsilon$.
SO(3)$\times$SO(N) fixed point}

As shown in Fig. \ref{symflow} for $N=3$, all fixed points with
symmetry $SO(4)\sim SO(3)\times SO(3)$ are unstable within a
$4-\epsilon$ expansion.  In contrast, as $N$ is increased a fixed
point with $SO(3)\times SO(N)$ symmetry appears that is fully
stable with respect to changes in the quartic interaction
parameters, including those perturbations that destroy
$SO(3)\times SO(N)$ symmetry. Such fixed point exists for $N\ge33$
and has
\begin{eqnarray}
v_1&=&-\frac{3\epsilon}{2K_d}h(N) \nonumber\\
v_2&=&\frac{\epsilon}{4K_d}\frac{1-72h(N)}{N+7} \nonumber\\
r_1&=&r_2=-2K_d\left\{(3N+2)v_1+(N+6)v_2\right\}
 \label{stableFixedPt}
\end{eqnarray}
where $u_1$, $u_2$ and $u_3$ are related to $v_1$ and $v_2$ by the
constraints (\ref{SO4Condition}), and the function
$h(N)=\left(N^2+8N-65+(N+7)\sqrt{N^2-34N+49}\right)^{-1}\sim
1/2N^2+O(1/N^3)$ is real only for $N>32$.

The two quadratic parameters $r_1$ and $r_2$ are relevant, tuning
the transition on a two dimensional phase diagram. The RG flow
equations (\ref{4-eRGequations}), linearized about the fixed point
(\ref{stableFixedPt}), yield two principal directions, $(\delta
r_1,\delta r_2)\propto(1,1)$ associated with the thermal exponent
$\lambda_t$, and $(\delta r_1,\delta r_2)\propto(-1,2)$ associated
with the anisotropy exponent $\lambda_g$.  From these we find the
critical exponents,
\begin{eqnarray}
1/\nu&=&\lambda_t=2-\epsilon
\left(1-\frac{10}{N}+\frac{28}{N^2}\right)+O(\frac{\epsilon}{N^3}),\nonumber\\
\phi&=&\lambda_g\nu=1-\epsilon
\left(\frac{3}{2N}+\frac{51}{2N^2}\right)+O(\frac{\epsilon}{N^3}).
\label{d4exponents}
\end{eqnarray}
Note that the crossover exponent
$\phi$ for the anisotropy is always less than one. This implies
\cite{Chaikin} that the phase boundaries meet as straight lines at
the critical point, and we find the same topology
of the phase diagram as shown in Fig.
\ref{figNearlySO4U2PositiveFluctuations}.

\subsection{Renormalization Group Analysis in $d=2+\epsilon$.
SO(3)$\times$SO(N) fixed point}

The runaway flows in the equations (\ref{4-eRGequations}) mean
that the system goes to strong coupling. In this limit the
magnitudes of the vectors ${\rm Re}\vec{\psi}$, ${\rm Im}\vec{\psi}$, and $\vec{N}$ have
already developed locally but the directions can still fluctuate on long length scales. For $N=3$
one of the runaway directions of equations (\ref{4-eRGequations})
corresponds to $u_3$, $v_2>0$ and $u_1$, $v_1<0$. The
corresponding strong coupling limit can be described by a triad of
vectors that are all mutually orthogonal.
\begin{eqnarray}
F=-\sum_{\langle xy \rangle} \left\{
K_1 \vec{e}_1(x)\cdot\vec{e}_1(y)
+K_2 ( \vec{e}_2(x)\cdot\vec{e}_2(y)+\vec{e}_3(x)\cdot\vec{e}_3(y) )
\right\}
\label{FTriad}
\end{eqnarray}
Here $\vec{e}_1$, $\vec{e}_2$, and $\vec{e}_3$ correspond to
$\vec{N}$, ${\rm Re}\vec{\Psi}$, and ${\rm Im} \vec{\Psi}$
respectively. The free energy (\ref{FTriad}) has
an explicit SO(2) charge
symmetry of rotations between $\vec{e}_2$ and $\vec{e}_3$.
The continuum version of this model is given by
\begin{eqnarray}
F = \int d^d x \left\{ \frac{1}{2g_1} (\nabla \vec{e}_1)^2+
\frac{1}{2g_2}((\nabla \vec{e}_2)^2+(\nabla \vec{e}_3)^2) \right\}
\label{FTriadContinuous}
\end{eqnarray}
Here $g_i \propto K_i^{-1}$ and the constraints $\vec{e}_i\cdot \vec{e}_j =
\delta_{ij}$ are implied.

Let us now discuss the phase diagram of (\ref{FTriad}) and
(\ref{FTriadContinuous}). For $K_{1,2} \rightarrow 0$ we
have a fully disordered phase. For $K_{1,2} \rightarrow
\infty$ we have a fully ordered phase that is a mixture of TSC and
AF. When $K_2=\infty$ the vectors  $\vec{e}_1$ and $\vec{e}_2$ are
ordered and there is an Ising type transition between TSC and
TSC+AF phases. For $K_1=\infty$ vector $\vec{N}$ is ordered and
there is an O(2) transition between the AF and TSC+AF states. For
$K_2=0$ there is a Heisenberg transition between the
disordered and the AF phases. For $K_3=0$ there is a transition
between the fully disordered and the TSC phases. What happens in
the interior of the phase diagram, however, is not clear.

\begin{figure}
\includegraphics[width=6.5cm]{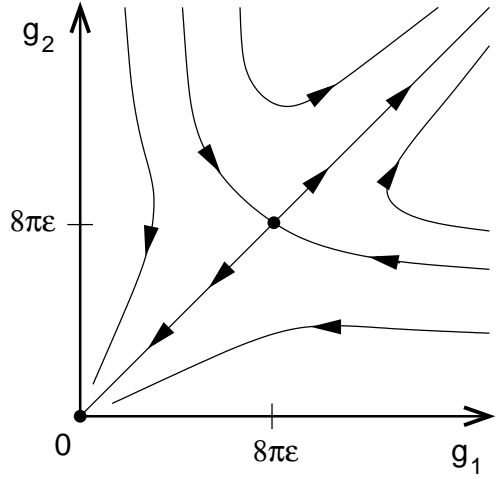}
\caption{Renormalization group flow of non-linear model eq.
(\ref{FTriadContinuous}), corresponding to equations
(\ref{N3d2eRGflow}). \label{N3d2efig}}
\end{figure}

When we apply the $d-2=\epsilon$ RG analysis to the model
(\ref{FTriadContinuous})
\cite{Friedan1985,Azaria1990,Kawamura1991,David1996,Kawamura1998}
we obtain the flow equations
\begin{eqnarray}
\frac{dg_1}{dl} &=& -\epsilon g_1
+\frac{g_1^2(g_2^2+g_1g_2-g_1^2)}{2\pi (g_1+g_2)^2}
\nonumber\\
\frac{dg_2}{dl} &=& -\epsilon g_2 +\frac{g_1^2g_2^2}{2\pi
(g_1+g_2)^2} \label{N3d2eRGflow}
\end{eqnarray}
The flow diagram is shown in fig. \ref{N3d2efig}.
We can see that it lacks the Ising and O(2) phase transitions.
This is not surprising, since the $d-2=\epsilon$ analysis works
well only for the spin-wave excitations of order parameters with
$N \geq 3$.

To shed some light on the phase diagram of (\ref{FTriad}) and
(\ref{FTriadContinuous}) we consider the large $N$ generalization
of this model. We assume that all vectors $\vec{e}_i$ have $N$
components
\begin{eqnarray}
F = \int d^d x \left\{ \frac{1}{2g_1} (\nabla \vec{e}_1)^2+
\frac{1}{2g_2}((\nabla \vec{e}_2)^2+(\nabla \vec{e}_3)^2)
+\frac{1}{g_3}((\vec{e}_1 \cdot\nabla \vec{e}_2)^2 +(\vec{e}_1 \cdot\nabla
\vec{e}_3)^2) +\frac{1}{g_4} (\vec{e}_2\cdot\nabla \vec{e}_3)^2
\right\} \label{FTriadContinuousLargeN}
\end{eqnarray}
The last two terms in equation (\ref{FTriadContinuousLargeN}) are
generated in the RG flow, even if they are absent in the
microscopic model (such terms are linearly independent of the first two
only for $N>3$).  The symmetry
breaking pattern of the non-linear model
(\ref{FTriadContinuousLargeN}) is
\begin{eqnarray}
{\rm O}(N)\times{\rm O}(2)/{\rm O}(N-3)\times {\rm O}(2)_{\rm diag}.
\end{eqnarray}
In order to
express the RG equations in a simple form we introduce the
variables $\eta_i$,
\begin{eqnarray}
\eta_1&=&\frac{1}{g_1},\nonumber\\
\eta_2&=&\frac{1}{g_2},\nonumber\\
\eta_3&=&\frac{1}{g_1}+\frac{1}{g_2}+\frac{2}{g_3},\nonumber\\
\eta_4&=&\frac{2}{g_2}+\frac{2}{g_4}. \label{g2eta}
\end{eqnarray}
These variables arise naturally in a matrix formulation of the
non-linear model Eq. (\ref{FTriadContinuousLargeN}), see Appendix
\ref{AppendixC}, where the RG calculation is outlined.
To one loop order we find the RG flow
\begin{eqnarray}
\frac{d\eta_1}{dl} &=& \epsilon \eta_1 -
\frac{1}{2\pi}\left(N-2+\frac{\eta_1^2-\eta_2^2-\eta_3^2}{\eta_2\eta_3}\right)
\nonumber\\
\frac{d\eta_2}{dl} &=& \epsilon \eta_2 -
\frac{1}{2\pi}\left(N-2+\frac{\eta_2^2-\eta_3^2-\eta_1^2}{2\eta_1\eta_3}
-\frac{\eta_4}{2\eta_2}\right)
\nonumber\\
\frac{d\eta_3}{dl} &=& \epsilon \eta_3 -
\frac{1}{2\pi}\left(N-2+\frac{N-3}{2}\frac{\eta_3^2-\eta_1^2-\eta_2^2}{\eta_1\eta_2}
-\frac{\eta_4}{2\eta_3}\right)
\nonumber\\
\frac{d\eta_4}{dl} &=& \epsilon \eta_4 -
\frac{1}{2\pi}\left(\frac{N-3}{2}\frac{\eta_4^2}{\eta_2^2}+\frac{\eta_4^2}{2\eta_3^2}\right)
\label{RG2+eLargeN}
\end{eqnarray}

The conditions $g_1=g_2$ and $g_3=g_4$ define a two dimensional
subspace over which the free energy (\ref{FTriadContinuousLargeN})
has the enhanced symmetry $SO(3)\times SO(N)$.  For arbitrary $N$,
the RG equations (\ref{RG2+eLargeN}) have an $SO(3)\times SO(N)$
fixed point
\begin{eqnarray}
\eta_1&=&\eta_2=(N-2-x)/\epsilon, \nonumber\\
\eta_3&=&\eta_4=x\eta_1, \label{d2eFixedPt}
\end{eqnarray}
where $x=(N-2+\sqrt{N^2-5N+5})/(N-1).$ Independent of $N$, this
point has one stable direction and one unstable direction within
the symmetric plane. The flow in directions
perpendicular to the symmetric plane depends on the value of $N$.
The case $N=4$ is special and is discussed in Appendix \ref{AppendixC}.
For all other $N\ge5$, the RG flow away from the
$SO(3)\times SO(N)$ plane has one stable and one unstable
direction.  Two relevant parameters are therefore necessary to
tune the transition, just as we found in the $d=4-\epsilon$
analysis for large $N$, eq. (\ref{stableFixedPt}).  Computing the
critical exponents $\nu$ and $\phi$, just as we did in the case
$d=4-\epsilon$, eq. (\ref{d4exponents}), we find,
\begin{eqnarray}
1/\nu&=&\epsilon, \nonumber\\
\phi&=&1-\frac{63}{16N^2}-O(\frac{1}{N^3}). \label{d2exponents}
\end{eqnarray}
Just as before, we find that for all finite $N$, $\phi$ is less
than one, leading to a phase diagram that is topologically
equivalent to that found in large $N$ expansion
and in $4-\epsilon$ RG analysis (see Fig.
\ref{figNearlySO4U2PositiveFluctuations}).
What's more, to leading order in $1/N$,
expansions about the upper and lower critical dimension, eqs.
(\ref{d4exponents}) and (\ref{d2exponents}), lead to the same
critical exponents,
\begin{eqnarray}
1/\nu&=&d-2+O(\frac{1}{N})\nonumber\\
\phi&=&1-O(\frac{1}{N}) \label{largeNexponents}
\end{eqnarray}
This supports the fact that for large $N$ the $SO(3)\times SO(N)$
fixed point changes adiabatically with dimension.

\subsection{Phase Diagram for $N=3$ in three dimensions}

We employed three approaches to study classical fluctuations in
systems with competing AF and non-unitary TSC: large $N$ expansion,
$d=4-\epsilon$ and $d=2+\epsilon$ RG analyses.  When $N$ is large all
three consistently predict a tetracritical point with enhanced
SO(3)$\times$SO(N) symmetry.  In the physically relevant case $N=3$
and $d=3$ the situation is less clear. For example, expansion from the
upper critical dimension ($d=4-\epsilon$) shows that such fixed point
appears only for $N \ge 33$. Expansion from the lower critical
dimension ($d=2+\epsilon$) gives an SO(3)$\times$SO(N) fixed point in
the RG flow for any $N\ge 3$, but these fixed points become tetracritical
points on the phase diagram only for $N \ge 5$.  It is possible that
in three dimensions even for $N=3$ there is a tetracritical
SO(3)$\times$SO(3) point.  The reason why perturbative expansions in
dimension fail to see it, is that they work well for small $\epsilon$
and extrapolations to $d=3$ should be treated with
caution\cite{Kawamura1998}.  Such scenario, however, would contradict
the results of classical Monte Carlo simulations in
Ref. \onlinecite{Diep1994}, in which the model (\ref{FTriad}) has been
analyzed for $K_1=K_2$. In that paper the bimodal distribution in the
energy histogram has been interpreted as a signature of the first
order transition.

\begin{figure}
\includegraphics[width=7.5cm]{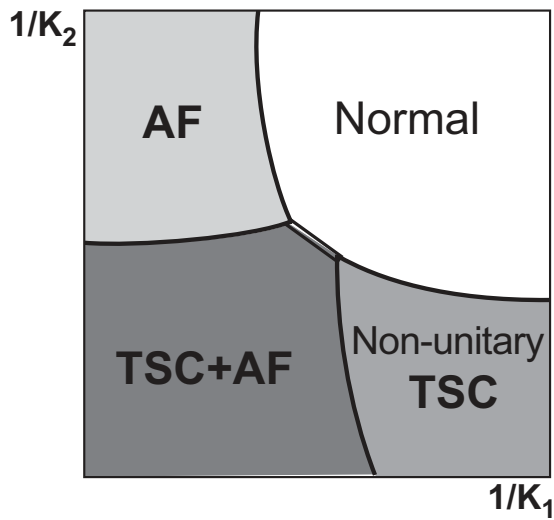}
\caption{Phase diagram for systems with competing AF and non-unitary
TSC orders described by the model (\ref{FTriad}) in three
dimensions. The tetracritical point in the mean-field phase diagram
 of the GL free energy in equation (\ref{NearlySO4GLenergy})
(see Fig. \ref{figNearlySO4U2Positive}) is replaced by a line of
direct first order transitions between a disordered and a mixed TSC/AF
phase.
\label{fig3}}
\end{figure}

The phase diagram that we propose for the model (\ref{FTriad}) and for
systems with competing AF and non-unitary TSC in general is shown in
Fig. \ref{fig3}. Thermal fluctuations turn a tetracritical point into
a line of direct first order transitions between a disordered and a
mixed TSC/AF phase. We expect, however, that the width of such first
order line is small and the transition is very weakly first order. We
conjecture that when approaching the transition between the normal and
the AF/TSC mixed phase, susceptibilities for the AF and TSC order
parameters start increasing as if dominated by the SO(3)$\times$SO(3)
tetracritical point. Only very close to the transition line the
divergencies are cut-off due to the transition being first order.

Finally, we note that eq. (\ref{FTriad}) with $K_1=0$ is among a
class of closely related models that have been studied
extensively in the context of frustrated magnetism\cite{mouhanna,
kawamura92,kunz93,kawamura85,mailhot94,bhattacharya94,
loison94,peles02,loison00b,loison99,itakura03}.  Numerical
studies of these models in $d=3$ dimensions yield non-universal
critical properties at the boundary of normal and TSC
phases\cite{kawamura92,kunz93,kawamura85,mailhot94,bhattacharya94,
loison94,peles02,loison00b,loison99}, and even evidence of a
first order transition\cite{itakura03}.  The non-perturbative
theoretical analysis of Ref. \onlinecite{tissier} supports the
latter scenario, claiming that the critical point observed
in non-linear sigma models in $d=2+\epsilon$ disappears at
$d_c=2.87$ in one such model,
being replaced by a weakly first
order transition above $d_c$.  The exact nature of the transition
seems to be very strongly model-dependent near $d=3$, and we
leave open the possibility that the transition between non-unitary
TSC and normal phases is weakly first order.

\section{Quantum SO(4) symmetry
}
\label{QuantumSection}


The microscopic system that motivated our discussion is an
assembly of Luttinger liquids weakly coupled in three dimensions.
It is useful to condense this system to a simpler effective
quantum model that concentrates on the low energy collective
degrees of freedom, such as AF and TSC order parameters and
rotations between them (such a description only applies in the
vicinity of the AF/TSC phase boundary shown in Fig.1).  Effective
quantum models have been discussed previously for spin systems
(see Ref \onlinecite{Sachdev_book} for a review), and systems with
singlet superconductivity competing either with charge density
wave order \cite{Yang1989,Yang1990,Zhang1990,Demler96}
or with antiferromagnetism
\cite{Zhang1997,Zhang1999,Markiewicz1999,Altman2002,Sachdev2003}.

A simple form for such an effective model is an SO(4)
quantum rotor model
\begin{eqnarray}
{\cal H}_{\rm r} &=& \frac{1}{2\chi_1} \sum_i \vec{S}_i^2 +
\frac{1}{2\chi_2} \sum_i \vec{I}_i^2
-J \sum_{\langle ij \rangle a\alpha} Q_{i,a\alpha}
Q_{j,a\alpha}\nonumber\\
&+& \tilde{u}_1 \sum_{iab\alpha\beta} Q^2_{i,a\alpha} Q^2_{i,b\beta}
+\tilde{u}_2 \sum_{iab\alpha\beta} Q_{i,a\alpha} Q_{i,a\beta}
Q_{i,b\alpha} Q_{i,b\beta}
\nonumber\\
&+&\delta r \sum_{i\alpha} (Q_{i,z\alpha}^2
- Q_{i,x\alpha}^2 -Q_{i,y\alpha}^2)
\label{QuantumRotor}
\end{eqnarray}
The model is obtained by coarse-graining the original lattice so that,
for a half-filled system, each site of the rotor model includes two
(or a larger even number, as necessary to include an integer number
of spin-triplet Cooper pairs) adjacent sites along the intrachain direction of the original lattice.
By combining the electronic operators that make up each rotor model site,
one can build three local spin and three local isospin operators
$\vec{S}_i$ and $\vec{I}_i$, and an SO(4) tensor order parameter
$Q_{i,a\alpha}$.  Following a procedure similar to Ref. \onlinecite{Altman2002},
one can show that the low energy properties of the system are given by the rotor commutation
relations,
\begin{eqnarray}
\left[ S_{i,\alpha},S_{j,\beta} \right] &=&
i \delta_{ij} \epsilon_{\alpha\beta\gamma}S_{i,\gamma}
\nonumber\\
\left[ I_{i,a},I_{j,b} \right] &=&
i \delta_{ij} \epsilon_{abc} I_{i,c}
\nonumber\\
\left[ S_{i,\alpha}, Q_{j,a\beta} \right] &=&
i \delta_{ij} \epsilon_{\alpha\beta\gamma}
Q_{j,a\gamma}
\nonumber\\
\left[ I_{i,a}, Q_{j,b\alpha} \right]
&=&
i \delta_{ij} \epsilon_{abc}
Q_{j,c\alpha}
\nonumber\\
\left[ Q_{i,a\alpha}, Q_{j,b\beta} \right]
&=& 0
\label{commutators}
\end{eqnarray}
These relations are analogous to
equations (\ref{SpinLuttinger}), (\ref{IsospinLuttinger}),
and (\ref{OrderParameterLuttinger}).
In equation (\ref{QuantumRotor})
the unit length constraint of the rigid rotor models is replaced by
the interaction terms $\tilde{u}_1$ and $\tilde{u}_2$. For $\delta r$
negative the system favors the AF state and for $\delta r$ positive the TSC
state is preferred. When $\delta r=0$ all generators of SO(4)
($\vec{I}=\sum_i \vec{I}_i$ and $\vec{S}=\sum_i \vec{S}_i$) commute
with the Hamiltonian (\ref{QuantumRotor}) and the system is exactly
SO(4) symmetric.

We can use equation (\ref{QuantumRotor}) to discuss
excitation spectra in various phases of the system.
We choose to orient the AF order parameter ($Q_{z\a}$)
in the $z$ direction so that $\langle Q_{zz}\rangle=N$.
Similarly we take $\langle Q_{xx}\rangle=\psi$ to describe unitary TSC,
and
$\langle Q_{xx}\rangle=\langle Q_{yy}\rangle=\psi$.
for non-unitary TSC.
With these choices we can linearize the equations of motion for the fluctuations
to obtain:
\begin{eqnarray}
\frac{dQ_{j,b\b}}{dt}&=&-\frac{1}{\chi_1}\sum_\a S_{j,\a}\e_{\a\b\ b}\av{Q_{bb}}
-\frac{1}{\chi_2}\sum_{\a} I_{j,\a}\e_{\a b \b}\av{Q_{\b\b}} \label{eqm1}\\
\frac{d S_{j,\a}}{dt}&=&\frac{J}{2}\sum_{\b',b'}\e_{\a \b' b'}\av{Q_{b'b'}}\sum_{\vec{\d}}
(Q_{j,b'\b'}-Q_{j,b'\b'+\vec{\d}})\label{eqm2}\\
\frac{d I_{j,a}}{dt}&=&\frac{J}{2}\sum_{\b',b'}\e_{a \b' b'}\av{Q_{\b'\b'}}\sum_{\vec{\d}}
(Q_{j,b'\b'}-Q_{j,b'\b'+\vec{\d}})
+4\delta r(\e_{az\b'}\av{Q_{\b'\b'}}Q_{j,z\b'}+\e_{azb'}Q_{j,b'z}\av{Q_{zz}})\label{eqm3}.
\end{eqnarray}

The above equations define a linear eigenvalue problem for
the frequencies of the collective modes and for the second quantized operators
\be
b\yd(\bk)=\sum_{c\ne\gamma}A^1_{c\gamma}Q_{c\gamma}(\bk)+\sum_a A^2_\a S_\a(\bk)+\sum_a A^3_a I_a(\bk),
\label{ayd}
\ee
which obey $\dot{b}\yd(\bk)=i[H,b\yd_\bk]=i\w_\bk b\yd(\bk)$.
The Fourier transforms of lattice operators are
defined by $\hat{O}(\bk)=N^{-1/2}\sum_j\hat{O}_je^{i\bk\cdot{\bf x}_j}$.
In the effective model, neutrons couple to the spin order parameter
$Q_{z\a}$, so that the low energy scattering intensity of polarized neutrons
is given by:
\begin{eqnarray}
\chi^{''}_\a(\bk + 2\bk_f,\omega) = \sum_{n} |\langle n | Q_{z\a}(\bk) | 0 \rangle |^2
\delta(\omega - \omega_{n0})
\label{chieff}
\end{eqnarray}
The momentum is shifted by $2k_f$ because a uniform $Q_{z\a}$ in (\ref{QuantumRotor})
corresponds
to a SDW order of momentum $2k_f$ in the microscopic model.
The weight associated with a particular collective mode created by $b\yd(\bk)$ is
\be
|\langle 0|b(\bk)Q_{z\a}(\bk)|0\rangle|^2\d(\w-\w_\bk)
=|\langle 0|[b(\bk),Q_{z\a}(\bk)]|0\rangle|^2\d(\w-\w_\bk)
\label{weight}
\ee
Here we used the fact that $b(\bk)$ annihilates the ground state.
The commutator can be calculated using (\ref{commutators}), once
the operator content of the mode $b(\bk)$ is determined.  We note that for
$\bk$ close to zero, neutrons couple to $S_\a$ instead of $Q_{z\a}$, in which
case,
\begin{eqnarray}
\chi^{''}_\a(\bk,\omega) = \sum_{n} |\langle n | S_{\a}(\bk) | 0 \rangle |^2
\delta(\omega - \omega_{n0}).
\label{chikzero}
\end{eqnarray}
However, in the following, we focus on neutron scattering near $2k_f$.

\begin{figure}
\includegraphics[width=6cm]{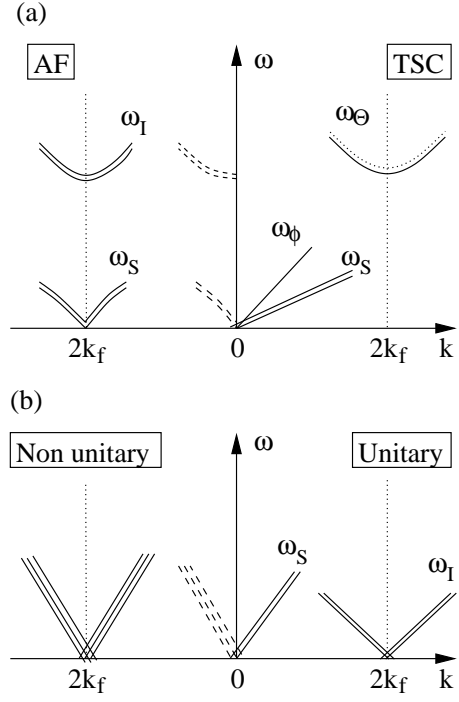}
\caption{Collective excitations in various phases. (a) In the AF
phase, the excitation spectrum consists of two massless spin waves
and two massive isospin waves. Due to translational symmetry
breaking, $2k_f$ is a reciprocal lattice vector, and these modes
also have non-zero weight at the dashed curves near $k=0$. The
spectrum in the TSC phase contains massless phase and spin modes,
as well as massive $\Theta$ modes. For unitary TSC, there is only
one such mode.  The non-unitary TSC (as well as the mixed phase)
contains a second, degenerate $\Theta$ mode, represented by the
dotted curve. (b) The SO(4) symmetric point ($\d r=0$). In a
unitary TSC this point corresponds to the transition between AF
and TSC. It is characterized by four gapless (Goldstone) modes (2
isospin and 2 spin). In the non-unitary case the SO(4) symmetric
point is inside the mixed AF/TSC phase. It supports only 3
(degenerate) Goldstone modes. This is because the order parameter
has a residual SO(3) symmetry as described in the text.
\label{Excitations}}
\end{figure}

The nature of collective excitations in the various phases is summarized in Figs.
\ref{Excitations}, \ref{fig:gap} and Table \ref{tab:all-par}.
We now provide a detailed analysis of the collective mode spectrum and the associated
neutron scattering intensity in each phase.
A complementary calculation of the neutron scattering intensity
of the $\Theta$-excitations in the unitary TSC, based on the microscopic model, is given in
Ref. \onlinecite{Podolsky2004}.

\subsection{Antiferromagnet}

In the AF phase $\av{Q_{zz}}=N$ and
all other order parameters vanish. Then Eqs. (\ref{eqm1}-\ref{eqm3})
decouple to four independent collective modes. The equations of motion for the pairs
$\{Q_{zx},S_y\}$ and  $\{Q_{zy},S_x\}$ yield the usual
AF spin waves with linear dispersion
reflecting broken SO(3) spin symmetry:
\be
\w_{AF,S}(\bk+2\bk_f)=N\sqrt{\frac{J z}{2\x_1}(1-\g_\bk)}\approx\sqrt{\frac{J}{2\chi_1}}~~|\bk|.
\label{afs}
\ee
Here $J_k=Jz/(2\x_1)(1-\g_\bk)$, $z$ is the lattice coordination
($z=6$ for a cubic lattice in three dimensions),
$\g_k=z^{-1}\sum_{\vec{\d}} e^{i\bk\vec{\d}}$, and $\vec{\d}$ are bond vectors.
Although $\bk$ and $\bk+2\bk_f$ are related by reciprocal lattice vectors in the
rotor model (\ref{QuantumRotor}), they are not in the microscopic Hamiltonian,
and the addition of $2\bk_f$ to the argument of (\ref{afs}) serves as a mnemonic for the
fact that the spin mode is centered primarily around $2\bk_f$.

Similarly the equations of motion for $\{Q_{xz},I_y\}$ and $\{Q_{yz},I_x\}$
describe two massive isospin waves:
\be
\w_{AF,I}(\bk+2\bk_f)=N\sqrt{\frac{4|\delta r|}{\chi_2} + \frac{J}{2\chi_2} k^2 }
\ee
These excitations correspond to rotations between the
AF and the TSC states and indicate proximity of the two ground
states. When $\delta r$ goes to zero, the mass of the isospin waves vanishes
reflecting an enhanced SO(4) symmetry. The isospin modes in the AF
phase do not couple to neutrons.

\subsection{Unitary triplet superconductor}

In the unitary TSC ($\delta r>0$, $\tilde{u}_2<0$) we choose
$\av{Q_{xx}}=\psi$, while all other order parameters vanish.
Eqs. (\ref{eqm1}) and (\ref{eqm2}) for the pairs $\{Q_{xy},S_z\}$ and
$\{Q_{xz},S_y\}$ yield two spin wave modes reflecting the broken spin symmetry:
\be
\w_{uTSC,S}(\bk)=\psi\sqrt{\frac{J}{2\chi_1}}~~|\bk|.
\ee
The creation operators for the two spin waves involve the generators $S_y$ and $S_z$
respectively. Substitution into (\ref{weight}) immediately shows that the
neutron scattering weight of these modes at $2\bk_f$ vanishes.

Eqs. (\ref{eqm1}) and (\ref{eqm3}) for the pair
$\{Q_{yx},I_z\}$ yields a gapless phase fluctuation mode reflecting
the broken charge U(1) symmetry in the TSC phase:
\be
\w_{uTSC,\f}(\bk)=\psi\sqrt{\frac{J}{2\chi_2}}~~|\bk|.
\ee
With the inclusion of Coulomb interactions, this mode becomes massive through
the Higgs phenomenon, with a mass of the order of the plasma frequency.
The creation operator of this isospin mode involves the generator $I_z$.
Substitution into (\ref{weight}) shows that it does not couple to $2k_f$
neutrons.

Finally the equations of motion for the pair $\{Q_{zx},I_y\}$
give the massive $\Theta$-mode:
\be
\w_{uTSC,\Theta}(\bk+2\bk_f)=\psi\sqrt{\frac{J}{2\chi_2}k^2+\frac{4\delta r}{\chi_2}}
\ee
The operator that creates this mode from the ground state is given by
\be
b\yd_\Theta(\bk)=\frac{1}{\sqrt{1+\x_2(Jk^2/2+4\delta r)}}
\left(\sqrt{\x_2(Jk^2/2+4\delta r)}Q_{zx}(\bk)+iI_y(\bk)\right)
\ee
When substituted into Eq. (\ref{weight}) it gives a neutron scattering intensity
at $\bk+2\bk_f$:
\be
|\langle 0|[b_\Theta(\bk),Q_{z\a}(\bk)]|0\rangle|^2\d(\w-\w_{uTSC,\Theta})
=\d_{\a x}\frac{\psi^2}
{1+\x_2(J k^2/2+4\delta r)}\d(\w-\w_{nTSC,\Theta}(\bk+2\bk_f))
\ee
As we approach the point $\d r=0$ with SO(4) symmetry, the gap of the $\Theta$-mode
vanishes as $\sqrt{\d r}$. Hence, the spectrum of the Hamiltonian
(\ref{QuantumRotor}) with $\tilde{u}_2<0$ is such that on both sides
of the AF/TSC transition we observe mode softening.  Mode
softening at the first order transition is a property of
the higher symmetry quantum critical points
\cite{Demler96,Zhang1997,Zhang1999}.
Exactly at the SO(4) symmetric point $\d r=0$ the system has
gapless spin and isospin wave doublets.

\subsection{Non unitary triplet superconductor}
For the case $\tilde{u}_2>0$ the TSC phase is non-unitary.
We choose $\langle Q_{xx} \rangle = \langle Q_{yy} \rangle=\psi$.
There is also a mixed phase where a non-vanishing AF order parameter $\av{Q_{zz}}=N$
appears in addition to the non unitary TSC order, considered below.
In the pure non-unitary TSC we find two spin wave modes $\{Q_{xz},S_y\}$
and $\{Q_{yz},S_x\}$ with linear dispersion:
\be
\w_{nTSC,S}(\bk)=\psi\sqrt{\frac{J}{2\x_1}}~~|\bk|
\ee
As in the unitary case, these spin waves do not couple to neutrons around $2\bk_f$.

The non unitary TSC also supports two degenerate
massive $\Theta$-modes $\{Q_{zx},I_y\}$ and $\{Q_{zy},I_x\}$ with dispersion
\be
\w_{nTSC,\Theta}(\bk+2\bk_f)=\psi\sqrt{\frac{J}{2\chi_2}k^2+\frac{4\d r}{\chi_2}}
\ee
These correspond to rotations of the real and
the imaginary parts of the TSC order parameter toward the AF. These excitations are
created by the operators:
\bea
b\yd_{\Theta_x}(\bk)&=&\frac{1}{\sqrt{1+\x_2(J k^2/2+4\d r)}}
\left(\sqrt{\x_2(Jk^2/2+4\d r)} Q_{zy}(\bk)-i I_x(\bk)\right)\nn\\
b\yd_{\Theta_y}(\bk)&=&\frac{1}{\sqrt{1+\x_2(J k^2/2+4\d r)}}
\left(\sqrt{\x_2(Jk^2/2+4\d r)} Q_{zx}(\bk)+i I_y(\bk)\right).
\eea
Substituting the $b_{\Theta_\a}$ operators in (\ref{weight}) we find the neutron scattering
weight near $2\bk_f$:
\bea
|\langle 0|[b_{\Theta_x}(\bk),Q_{z\a}(\bk)]|0\rangle|^2
\d(\w-\w_{nTSC,\Theta})&=&\d_{\a y}\frac{\psi^2}{1+\x_2(J k^2/2+4\d r)}\d(\w-\w_{nTSC,\Theta}(\bk+2\bk_f))\nn\\
|\langle 0|[b_{\Theta_y}(\bk),Q_{z\a}(\bk)]|0\rangle|^2
\d(\w-\w_{nTSC,\Theta})&=&\d_{\a x}\frac{\psi^2}{1+\x_2(J k^2/2+4\d r)}\d(\w-\w_{nTSC,\Theta}(\bk+2\bk_f))
\eea

The phase fluctuation mode in the non unitary TSC phase differs from
its counterpart in the unitary case. The Eqs. (\ref{eqm1}-\ref{eqm3}) for
$\{Q_{xy},Q_{yx},S_z,I_z\}$ cannot be decoupled, giving a mode that involves both
spin and isospin generators. Due to the residual symmetry generated by $S_z+I_z$,
the mode with $Q_{xy}-Q_{yx}$, drops out of the spectrum.
The remaining excitation follows the dispersion
\begin{eqnarray}
\w_{nTSC,\f}(\bk)=\psi\sqrt{J\left(\frac{1}{\chi_1}+\frac{1}{\chi_2}\right)}~~k.
\label{phasenuTSC}
\end{eqnarray}
As in the unitary case, Coulomb interactions make this a massive mode, with a mass
of the order of the plasma energy.
The creation operator of this mode involves the generators $I_z$ and $S_z$.
Substitution into (\ref{weight}) gives vanishing neutron scattering intensity
at $2\bk_f$.

\subsection{Mixed phase}

\begin{figure}
\includegraphics[width=12cm]{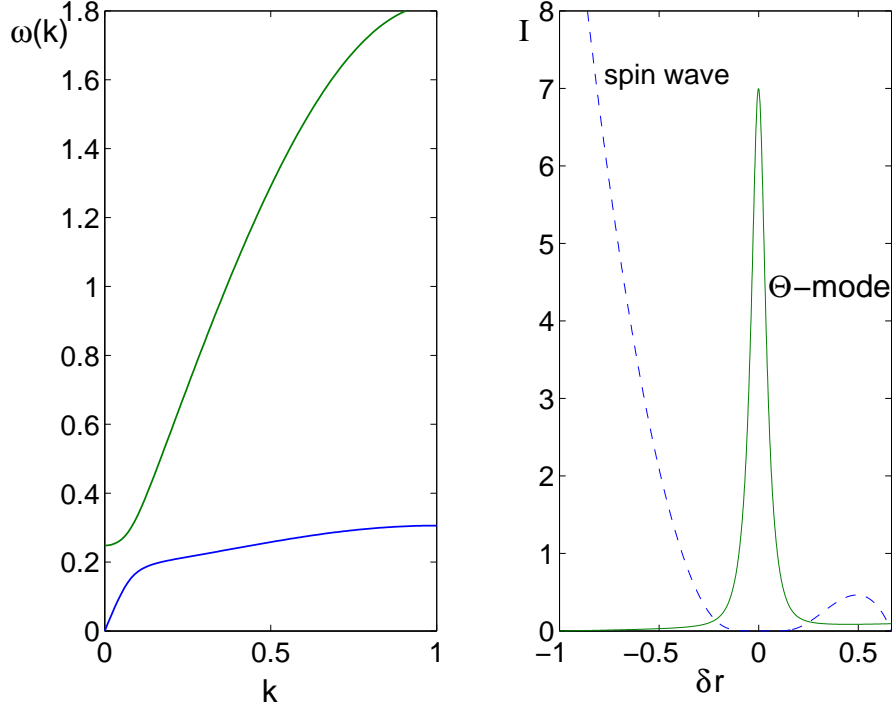}
\caption{Spin waves and $\Theta$-modes in the mixed phase. (a)
Dispersions ($\d r>0$) of the mode clearly show mixing between the
$\Theta$-mode and the spin wave. Each of the shown modes is doubly
degenerate. (b) Neutron scattering intensity of the modes at
wave-vector $k=2k_f+0.1\pi$ as a function of the tuning parameter
$\d r$. Note that the weight of the spin wave modes goes to zero
at the SO(4) symmetric point $\d r=0$. $\Theta$-modes are strongly
enhanced near $\d r=0$.} \label{fig:wmixed}
\end{figure}

\begin{figure}
\includegraphics[width=12cm]{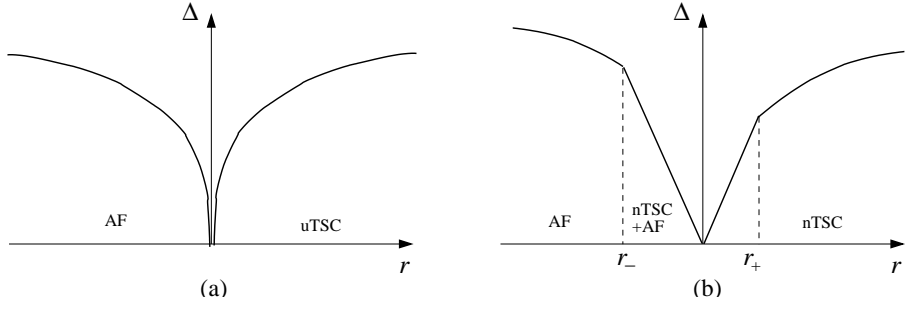}
\caption{Gap of the $\Theta$ modes softens toward $\d r=0$
reflecting the enhanced SO(4) symmetry at that point. (a) The gap decreases as $\sqrt{|\d r|}$
in the case $\tilde{u}_2<0$ (unitary TSC).
(b) In the case $\tilde{u}_2>0$ there is a change from $\sqrt{|\d r|}$ behavior
in the pure phases to linear decrease at smaller $|\d r|$, inside the mixed non-unitary TSC and AF
phase.}
\label{fig:gap}
\end{figure}

In the AF/non-unitary TSC mixed phase, the order parameters form an orthogonal triad
$\vec{\psi_1}=\psi \hat{x},~ \vec{\psi_2}=\psi \hat{y}, ~\vec{N}=N\hat{z}$, i.e.
$\av{Q_{xx}}=\av{Q_{yy}}=\psi$ and $\av{Q_{zz}}=N$.
It is easy to verify that the phase fluctuation mode remains unchanged,
its dispersion given by Eq. (\ref{phasenuTSC}).
However, other modes are complicated due to the fact that
Eqs. (\ref{eqm1}-\ref{eqm3}) couple the coordinates
$\{Q_{xz},Q_{zx},S_y,I_y\}$ and similarly $\{Q_{yz},Q_{zy},S_x,I_x\}$.
Solution of the eigenvalue equations yields two collective modes
for each of the above coordinate sets.
One is a massive ``$\Theta$" mode
and the other a gapless spin wave-like mode:
\be
\w_{S,\Theta}=\sqrt{\left(\frac{2\d r\phi}{\x_2}
+\frac{J_\bk\rho}{2\x_t}\right)\mp\sqrt{\left(\frac{2\d r\phi}{\x_2}
+\frac{J_\bk\rho}{2\x_t}\right)^2-\frac{J_\bk}{\x_1\x_2}(J_\bk \phi^2+4\d r\phi\rho)}}
\label{wmixed}
\ee
Where $\x^{-1}_t\equiv \x^{-1}_1+\x^{-1}_2$, $J_\bk\equiv Jz/2(1-\g_\bk)$,
$\phi\equiv\psi^2-N^2$, and $\rho\equiv \psi^2+N^2$.
To calculate the spectrum in the
mixed phase as a function of the tuning parameter $\d r$,
we find the values of the order parameters at a given $\d r$
from the mean field theory of (\ref{QuantumRotor}). Specifically we use the
result $\psi^2-N^2=\d r/\tilde{u}_2$. Fig. \ref{fig:wmixed}(a)
gives an example of the dispersions obtained for a particular value of $\d r$
within the mixed phase. The asymptotic form of the excitation energies
at small wave vectors is given by:
\bea
\w_{S}&\sim&\sqrt{\frac{J\rho}{2\x_1}}~|\bk|\nn\\
\w_{\Theta}&\sim&\sqrt{\frac{J\rho}{2\x_2}k^2+\frac{4 \d r^2}{\tilde{u}_2\x_2}}.
\eea
Fig. \ref{fig:wmixed}, demonstrates that the exact dispersion (\ref{wmixed}) deviates
from these asymptotic forms already at relatively small wave vectors. This is due to
the strong mixing between spin and isospin modes.
Due to this mixing, both spin waves
and $\Theta$-modes carry some weight in the neutron scattering
intensity (see Fig. \ref{fig:wmixed}(b)).
The scattering intensity associated with the spin wave mode vanishes
in the vicinity of $\d r=0$. On the other hand the intensity of the $\Theta$ modes becomes
dramatically enhanced. Another unique feature of the phase with mixed
non-unitary TSC and AF order is a linear with $|\d r|$ softening of the
$\Theta$-excitation gap. Compare this to the $\sqrt{\d r}$ softening in the unitary TSC
(See also Fig. \ref{fig:gap}).

The SO(4) symmetric point $\d r=0$ needs special consideration.
Here $N^2=\psi^2$ and the order parameter is
invariant under the SO(3) group generated by ${\bf I}+{\bf S}$.
This implies that there are only three Goldstone modes at this point. Indeed, a direct calculation
at the SO(4) symmetric point gives three degenerate modes with dispersion:
\be
\w_{nSO(4)}(k)=\psi\sqrt{\frac{J}{2}\left(\frac{1}{\chi_1}+\frac{1}{\chi_2}\right)}~|\bk|
\ee

Note that the number of Goldstone modes at the SO(4) point is
different in a unitary and non-unitary TSC. In the unitary case the
spin and isospin SO(3) symmetries are broken separately with a residual $U(1)\times U(1)$
symmetry of the order parameter.
local gauge freedom associated with each. This leads
to four Goldstone modes. In the non unitary case, on the other hand,
there is residual SO(3) symmetry of the order parameter, corresponding
to ${\bf I}+{\bf S}$ rotations as discussed
above. Consequently, this system has only three Goldstone modes.

The gapless spin waves and phase modes that we found away from the SO(4) symmetry
are generic to
systems that break spin SO(3) and charge U(1) symmetries.  However,
$\Theta$ excitations, which can be thought of as massive isospin waves,
are not. Their presence shows the proximity of AF and TSC phases and
their softening at the point $\d r=0$ should provide a unique signature of
the SO(4) symmetry of the system.
\begin{table}[h]
\begin{center}
\begin{tabular}{|c||c|c|c|c|}
\hline
Phase  & Order parameter & Residual symmetry & Goldstone (massless) modes &
Pseudo-Goldstone (massive) modes \\ \hline\hline
AF & $\av{Q_{zz}}$ &  U(1)$\times$U(1)~~$(S_z,I_z)$ & 2~~ $(S_x,S_y)$ & 2~~ $(I_x,I_y)$ \\
uTSC & $\av{Q_{xx}}$ & U(1)~~$(S_x)$ & 3~~$(S_z,S_y,I_z)$ & 1~~ $(I_y)$\\
nuTSC & $\av{Q_{xx}}=\av{Q_{yy}}$ & U(1)~~$(S_z+I_z)$ & 3~~$(S_z,S_y,S_z-I_z)$ & 2~~$(I_y,I_x)$\\
nTSC+AF & $\av{Q_{xx}}=\av{Q_{yy}}$,$\av{Q_{zz}}$
& U(1)~~$(S_z+I_z)$ & 3~~$(S_z,S_y,S_z-I_z)$ & 2~~$(I_y,I_x)$\\
unitary SO(4) & $\av{Q_{xx}}$ & U(1)$\times$U(1)~~$(I_x,S_x)$ & 4~~$(S_z,S_y,I_z,I_y)$ & 0 \\
non unitary SO(4) & $\av{Q_{xx}}=\av{Q_{yy}}=\av{Q_{zz}}$ & SO(3)~~$({\bf I}+{\bf S})$ &
3~~$ ({\bf I}-{\bf S})$ & 0\\\hline
\end{tabular}
\vskip0.4pc \caption{Symmetry breaking and collective modes.
Here uTSC and nuTSC stand for unitary and non-unitary TSC respectively,
nTSC+AF corresponds to a mixed phase of non-unitary TSC and antiferromagnetism
away from the SO(4) symmetric point.
} \label{tab:all-par}
\end{center}
\end{table}

\section{SO(4) symmetry in a strongly anisotropic Fermi liquid}
\label{SectionInterchain}

Thus far in the analysis we have treated the case of weakly
coupled Luttinger liquids, where we showed that SO(4) symmetry
describes the phase diagram and collective modes of a system near
the AF/TSC phase transition.  However, there is a tendency away
from Luttinger behavior as pressure is increased towards the
superconducting state, as supported by the observation of
field-induced SDW phases at high magnetic
fields\cite{Kwak1982,Bando1982,Brusetti1982b,Gorkov1984,Dupuis1993,Yakovenko1996},
by optical measurements\cite{Vescoli1998}, and by low temperature
transport experiments\cite{Dressel2004}. A review of the normal
state of Bechgaard salts at low magnetic fields is given in
Ref.~\onlinecite{Bourbonnais1999}. In this section, we will
consider the effects of interchain hopping in the extreme case
where it is large enough to destroy all remnants of Luttinger
liquid physics, and make the system into a highly anisotropic
Fermi liquid instead.  We will see that even in this limit,
despite the loss of nesting, an approximate SO(4) symmetry
remains.

We begin by looking at the classical SO(4) symmetry of the GL free
energy.  In order to investigate this symmetry, it is sufficient
to consider the quartic GL terms.  This follows from the fact that
our analysis of the phase diagram includes explicitly anisotropy
in the quadratic terms, see eq. (\ref{NearlySO4GLenergy}), which
is tuned to zero by pressure at the AF/TSC transition.  As shown
in Appendix~\ref{AppendixGLDerivation}, a microscopic derivation
of the GL parameters starting from a weakly interacting Fermi
liquid leads to the following form for the quartic GL terms,
\begin{eqnarray}
F_4= A \left( 2(|\vec{\Psi}|^2)^2 -
       |\vec{\Psi}^2|^2 \right)
+ B (\vec{N}^2)^2 + 2(C+D)|\vec{\Psi}|^2\vec{N}^2 -4D
|\vec{\Psi}\cdot\vec{N}|^2, \label{GLfromFL}
\end{eqnarray}
where, for a perfectly nested Fermi surface,
$A=B=C/2=D=\frac{7\zeta(3)}{16\pi^2v_f T^2}$ satisfy the SO(4)
symmetry conditions~(\ref{SO4Condition})
\begin{eqnarray}
A&=&D,\nn\\
B&=&D,\nn\\
C/2&=&D,\nn
\end{eqnarray}
and the sign of the coefficients corresponds to the unitary TSC
case. In the presence of interchain coupling $t_b$, the single
electron spectrum becomes
\begin{eqnarray}
\xi_\bk=-2t_a\cos k_a-2 t_b\cos k_b-\mu\label{singlePart}.
\end{eqnarray}
Although Bechgaard salts are triclinic, and expression
(\ref{singlePart}) applies to rectangular lattices only, it gives
a good approximation to the low energy quasiparticle states of the
system. Here, we take $t_a=250$ meV, $t_b=20$ meV, as estimated
from plasma frequency measurements\cite{Jacobsen1983} and band
structure calculations\cite{Grant1983}.  In addition, we take
$\mu=\sqrt{2} t_a$, corresponding to a quarter-filled band and a
nesting vector $\bQ=(2k_f,\pi)\approx (\pi/2,\pi)$. We note that,
in the Fermi liquid description, dimerization only affects very
high energy quasiparticles, and we exclude it from
(\ref{singlePart}).

\begin{figure}
\includegraphics[width=10cm]{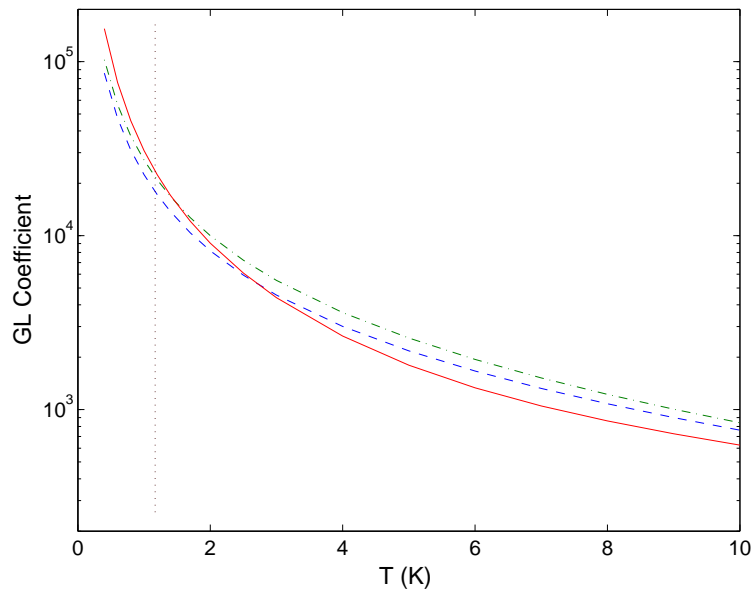}
\caption{Quartic Ginzburg-Landau coefficients (\ref{GLfromFL}) for
a strongly anisotropic Fermi liquid, for the choice of parameters
$t_a=250$ meV, $t_b=20$ meV.  The solid curve shows $\sqrt{AB}$,
the dashed curve $D$, and the dash-dotted curve $(C+D)/3$.
Although the three curves do not coincide, as would be required by
SO(4) symmetry, they trace similar trajectories all the way down
to the critical temperature $T_c=1.2$ K (vertical dotted line). At
temperatures higher than shown, the three curves converge as
nesting is restored. \label{fig:interchainGL}}
\end{figure}

At first glance, interchain hopping seems to have a devastating
effect on the SO(4) symmetry.  The nesting vector $\bQ$ no longer
connects the right and left moving Fermi surfaces exactly. Hence,
while the coefficient $A$ is insensitive to $t_b$, the low
temperature divergence in the coefficients $B$, $C$, and $D$ is
preempted by the loss of nesting.  Instead, these coefficients
saturate at a temperature of the order of $t_b^2/t_a\approx 20$ K,
changing the ratio $A/B$ from unity at high temperatures to about
$10$ at $T_c=1.2$ K. However, nesting strongly affects
antiferromagnetism only, and not superconductivity.  Hence, it's
effects on the GL parameters grow in proportion to the number of
times that each GL parameter multiplies $\vec{N}$ in
(\ref{GLfromFL}).  Thus, most of the effect can be absorbed into
the normalization of the field $\vec{N}$. While the fields
$\vec{N}$ and $\vec{\Psi}$ cannot be normalized independently in
the full GL free energy, as this would change the ratio of
gradient terms $\half (|\nabla \vec{\Psi}|^2+(\nabla \vec{N})^2)$,
such scaling is allowed when considering the mean-field properties
of the system. The conditions for SO(4) symmetry at mean-field
level are then
\begin{eqnarray}
\sqrt{AB}&=&D,\nn\\
(C+D)&=&3D.\label{SO4ConditionFL}
\end{eqnarray}
Thus, at the mean-field level, SO(4) symmetry is only broken
weakly. This is illustrated in Fig.~\ref{fig:interchainGL}, where
the left and right hand sides of the conditions
(\ref{SO4ConditionFL}) are evaluated explicitly. Despite the
strong variation in the values of the different GL coefficients,
the curves shown in Fig.~\ref{fig:interchainGL} trace similar
trajectories, indicating the approximate SO(4) symmetry.  At
$T_c=1.2$ K, we find $A=7.0\times 10^4$, $B=7.3\times 10^3$,
$C=3.1\times 10^4$, and $D=2.1\times 10^4$, leading to
$\sqrt{AB}/D=1.06$ and $(C+D)/(3D)=0.82$ ($\vec{N}$ was rescaled
by a factor of 1.76). Thus, the conditions (\ref{SO4ConditionFL})
deviate from exact SO(4) symmetry by less than 20\% at $T_c=1.2
K$.

While classical SO(4) symmetry is no longer exact, the phase
diagram derived in previous sections does not change in important
ways. At the mean-field level, the 20\% variation from the SO(4)
conditions (\ref{SO4ConditionFL}) can open a very narrow mixed
phase between the TSC and AF phases.   Hence, as pressure is
varied from the AF phase to the TSC phase, there will no longer be
a discontinuity in the order parameters or in the density.
Instead, these quantities will show a smooth but very rapid
variation as pressure transverses the mixed phase.  Therefore, due
to the narrowness of the mixed phase, the system will be very
sensitive to disorder. For realistic systems, which have
impurities and crystal defects, the mixed phase will segregate
into inhomogeneous regions of AF and TSC, just as found in the
coexistence phase in the strictly first order case. However,
unlike the case of a first order transition, the inhomogeneous
behavior will be apparent even when the phase diagram is tuned by
experimental variables that are intrinsic. We note that the
narrowness of the mixed phase, and the corresponding sensitivity
to disorder, is a direct consequence of the proximity of the
system to SO(4) symmetry.

We now consider the fluctuation-induced first order transition
between the AF and normal phases near the SO(4) symmetric point.
We note that in order to alter the topology of the phase diagram,
the bare GL parameters must differ enough from the SO(4) symmetric
values to divert the RG flow near a new critical point. While
symmetry can play an important role in an RG flow, it is difficult
to conceive of a situation where {\it reduction} of symmetry would
lead to softening of the first order transitions into second
order.  Thus, we expect the first order transition between AF and
normal phases discussed earlier to still be present. Finally, we
note that the case we consider in this section is extreme, in that
we look study the system as a weakly interacting, strongly
anisotropic Fermi liquid.  This probably gives a strong
overestimate of the magnitude of the breaking of classical SO(4)
symmetry in real systems, which are likely to lie between the
Fermi liquid limit and the weakly coupled Luttinger liquid limit,
where SO(4) is a good symmetry.

Before concluding this section, we briefly discuss the effects of
interchain coupling on the quantum SO(4) symmetry. One can look
for this symmetry by verifying the existence of the $\Theta$
resonance in a strongly anisotropic Fermi liquid formalism. Inside
the TSC phase, this can be done using RPA type calculations, which
include the AF particle-hole and $\Theta$ resonance
particle-particle channels (see e.g. Ref.~\onlinecite{Demler98}).
Results of these calculations will be reported elsewhere.  The
main effect of interchain hopping is to fix the transverse
components of the nesting vector to $\bQ=(2k_f,\pi,\pi)$.  Thus,
the $\Theta$ resonance for quasi-one dimensional systems is a
collective mode, whose quantum numbers are spin zero, charge two,
and wave vector $\bQ$. The presence of interchain coupling also
introduces broadening of the $\Theta$ mode, and prevents it from
softening all the way down to zero energy at the AF/TSC phase
transition. Instead, we expect the minimum energy of the
excitation to be of the order of $t_b^2/t_a=16$ meV.

\section{Experimental signatures of the SO(4) symmetry}
\label{SectionSignatures}

The interplay of AF and SC in the organic material
(TMTSF)$_2$PF$_6$ has been a subject of active investigation
\cite{Jerome1980,Andres1980,Takahashi1989}. There is strong
experimental evidence supporting that superconducting order is
spin triplet, as discussed in Section \ref{SectionIntro}. In
addition, (TMTSF)$_2$PF$_6$ has a quasi one-dimensional structure,
as the anisotropy of electron tunneling along the chains (a), in
the planes (b), and perpendicular to the planes (c) is of the
order of $t_a:t_b:t_c=250:25:1$.  Hence, (TMTSF)$_2$PF$_6$ is a
good candidate for comparison with the theoretical model discussed
in this paper. Following the discussion in Section
\ref{SectionGLDiscussion} and Appendix \ref{AppendixGLDerivation}
we expect the triplet order parameter in this material to be
unitary. The phase diagram for this case was obtained in Section
\ref{SectionUnitaryTSC}. In Ref. \onlinecite{Podolsky2004} we
compare this to the experimental phase diagram of
(TMTSF)$_2$PF$_6$\cite{Vuletic2002,Kornilov2003}. One consequence
of having enhanced symmetry at a phase transition is the
suppression of the critical temperature due to fluctuations of one
order parameter into the other.  This may contribute to the
drastic drop in $T_{\rm AF}$ as pressure is increased near the
AF/TSC phase boundary in Bechgaard salts\cite{Vuletic2002}.

The first order transition between AF and TSC phases near the
critical point, Fig. \ref{figLargeNU2Negative}, leads to a regime
of frustrated phase separation, with domains of one phase inside
the other.  The volume fractions of each phase are governed by the
Maxwell construction, while the size of individual domains is
determined by the competition between short-range and long-range
parts of the Coulomb interaction\cite{Carlson2002}.  If the
domains are distributed randomly, the total resistance of the
system may be found using an effective medium approximation.  This
implies, for example, that the system is superconducting when the
TSC phase is beyond the percolating threshold.  On the other hand,
it is possible that the TSC domains are not distributed uniformly
in the system, and are more favorable on the surface of the
sample.  In this case, the TSC regions can ``short-circuit" the
system even before they reach the percolation condition for the
bulk.  Transport properties consistent with this scenario of an
inhomogeneous system have been reported in
Ref.~\onlinecite{Vuletic2002}.

\begin{figure}
\includegraphics[width=6cm]{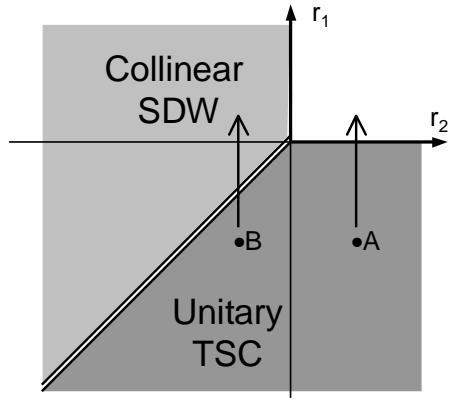}
\caption{The effect of a magnetic field on the superconducting
state.  For points (A) far from the AF/TSC boundary, the magnetic
field destroys superconductivity leading to a normal state. For
points (B) close to the boundary, a magnetic phase is stabilized
instead.  The double line denotes a first order transition, which
expands into a AF/TSC coexistence region in the experimental phase
diagram.  Here we focus on the unitary case, but similar effects
can be seen for the non-unitary case near the AF/TSC mixed phase.
Fields of the order of 100 mT are sufficient for a significant
enhancement in antiferromagnetism to be observed. This is in
contrast with field-induced SDW phases, which require fields in
excess of 5 T.\label{Bfield}}
\end{figure}

An interesting direction for exploring competition between AF and
TSC phases is to use magnetic field experiments in the
superconducting state near the AF/TSC phase boundary.  Magnetic
field produces orbital currents that strongly suppress electron
pairing and leads to a formation of an Abrikosov vortex lattice.
Suppression of the AF order by Zeeman effect is much smaller.
Thus, we expect magnetic fluctuations to become strongly enhanced
in the mixed
state\cite{Zhang1997,Arovas1997,Demler2001,Zhang2002,Kang2003}.
Since the critical field along the c-axis is $H^c_{c2}= 100
mT$\cite{Lee1997}, an applied magnetic field along the c-axis on
the order of a few mT can have a strong effect, see
Fig.~\ref{Bfield}. This is in contrast with effects such as field
induced SDW's and reentrant superconductivity, which require
fields of at least 5 T for their
observation\cite{Ishiguro1998,Kwak1982,Bando1982,Brusetti1982b,Gorkov1984,Dupuis1993,Yakovenko1996}.
For pressures close to the AF/TSC phase boundary and for slightly
larger magnetic fields there may also be a quantum phase
transition in which long range AF order develops inside the vortex
phase.   We note that strong sensitivity of $1/T_1$ to magnetic
fields in the superconducting state of (TMTSF)$_2$PF$_6$ have been
reported in Ref.~\onlinecite{Lee2000}.  Here, increasing the
magnetic field along the b-axis from 12.8 mT to 232 mT results in
a large increase of $1/T_1$, consistent with the enhancement of
antiferromagnetism that we propose. Earlier specific heat
measurements in Ref.~\onlinecite{Garoche1982} already showed that
when the superconducting order in (TMTSF)$_2$ClO$_4$ is suppressed
by a magnetic field, the system goes into a semimetallic state
with a suppressed quasiparticle density of states. This is
consistent with developing AF order, thus opening a gap in the
quasi particle spectrum. It may be interesting to study further
the enhancement of magnetic order in the mixed state with neutron
scattering \cite{Katano2000,Lake2001,Lake2002,Khaykovich2002}, NMR
\cite{Curro,Mitrovic,Kakuyanagi}, and $\mu$SR\cite{Miller}
experiments.

In Ref.~\onlinecite{Podolsky2004} (see also section
\ref{QuantumSection} of this paper) we discuss that direct
observation of the $\Theta$ mode in the superconducting phase
should be possible through neutron scattering.  The most important
feature of the $\Theta$-resonance, which identifies it as a
generator of the SO(4) symmetry, is the pressure dependence of the
resonance energy inside the TSC phase. When the pressure is
reduced and the system is brought toward the phase boundary with
the AF phase, we predict the energy of the $\Theta$-resonance to
be dramatically decreased. Mode softening is not expected
generically at first order phase transitions and provides a unique
signature of the SO(4) quantum symmetry.  We note that due to
interchain hopping, the center of mass momentum of the $\Theta$
excitation in quasi-one dimensional systems is $(2k_f,\pi,\pi)$.

\begin{figure}
\includegraphics[width=4cm]{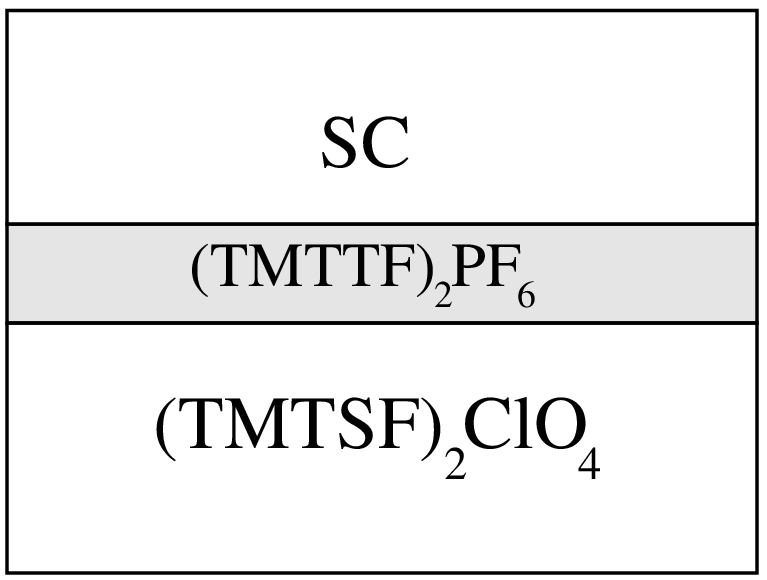}
\caption{Tunneling experiment for detecting the $\Theta$
excitation in (TMTSF)$_2$ClO$_4$ material. A singlet
superconducting material with a higher transition temperature than
(TMTSF)$_2$ClO$_6$ provides a reservoir of Cooper pairs that can
couple resonantly to $\Theta$ pairs. Momentum mismatch between the
Cooper pairs in SC and $\Theta$ pairs in (TMTSF)$_2$ClO$_4$ is
compensated by scattering of electrons in a layer of the SP
material (TMTTF)$_2$PF$_6$. \label{TunnelingExperiment}}
\end{figure}

Another approach to detect the $\Theta$ excitation involves
tunneling experiments with the SSC/(TMTSF)$_2$ClO$_4$ junction
shown in Fig. \ref{TunnelingExperiment} (analogous experiments in
the context of $\pi$ excitations in the high Tc cuprates are
discussed in Ref. \onlinecite{Bazaliy1997}).  A singlet
superconductor provides a reservoir of Cooper pairs that can
couple to $\Theta$ pairs in (TMTSF)$_2$ClO$_4$.  One needs to
overcome, however, the momentum mismatch between the two types of
pairs. A possible approach is to use an intermediate layer of the
quasi 1d material (TMTTF)$_2$PF$_6$. This salt is quarter filled
and displays spin-Peierls (SP) order. The modulations of the SP
order thus have a periodicity of four TMTTF sites, matching the
$(2k_f,\pi,\pi)$ wave vector of (TMTSF)$_2$ClO$_4$. The small
mismatch between the two wave vectors, due to differences in the
lattice constant in these compounds, can be compensated by a
parallel magnetic field \cite{Scalapino1970}. We expect peaks in
the current-voltage characteristics of the junction when the
voltage bias compensates the energy difference between Cooper and
$\Theta$ pairs
\begin{eqnarray}
2eV = \omega_\Theta
\end{eqnarray}
Peaks in $IV$ should be present even above the superconducting
transition temperature of (TMTSF)$_2$ClO$_4$ and only require the
other material to be superconducting. The choice of
(TMTSF)$_2$ClO$_4$ is made as this material is likely to be close
to the AF/TSC transition at ambient pressure
\cite{Bourbonnais1984}. This eliminates the need for pressure
cells, which would make the experiments much more difficult.

\section{Summary}
\label{SectionSummary}

The primary purpose of this paper has been to discuss the
competition of antiferromagnetism and triplet superconductivity in
quasi one-dimensional systems, such as Bechgaard salts
(TMTSF)$_2$X.  The point of departure of our work is the existence
of enhanced symmetry, that unifies the two order parameters, in
one dimensional systems of interacting electrons. Analysis of the
Luttinger liquid model presented in Section \ref{LuttingerSection}
showed that the usual charge and spin U(1)$\times$SO(3) symmetry
is enhanced to a higher SO(3)$\times$SO(4) symmetry on the
transition line between the two phases for incommensurate band
filling.  For half-filled systems and weak umklapp scattering, the
enhanced symmetry group becomes SO(4). Weak coupling between
chains, that enables true long range order, is expected to perturb
the SO(4) symmetry only slightly.

In sections \ref{SectionGLDiscussion}, \ref{SectionThermal},
\ref{SectionUnitaryTSC}, and \ref{SectionNonUnitary} we studied
the finite temperature phase diagram for systems with SO(4)
symmetry. For the unitary case, a mean field analysis shows that
SO(4) symmetry requires a direct first order transition between
TSC and AF phases. In addition, fluctuations of the order
parameters turn a portion of the boundary between AF and normal
phases into a first order transition and also lead to a weakly
first order transition between TSC and normal phases.  For the
non-unitary case, SO(4) symmetry requires two second order
transitions between TSC and AF phases.  We find that the system is
close to having an SO(4) symmetric tetracritical point, but
thermal fluctuations stretch this point into a short line of
direct first order transitions from the normal state to the mixed
state.  Our results have direct implications for quasi
one-dimensional organic superconductors from the (TMTSF)$_2$X
family, which are likely to be unitary triplet superconductors.
For example, first order transitions between the AF and the  TSC
phases, and between the Normal and the AF phases explain the
AF/TSC and the AF/Normal coexistence regions found in the phase
diagram of (TMTSF)$_2$PF$_6$.

In section \ref{QuantumSection} we analyze collective excitations
in various phases and demonstrate that SO(4) leads to the
existence of a new collective mode, the $\Theta$ excitation, which
describes rotations between the AF and the TSC phases.  In section
\ref{SectionSignatures} we study possible experimental tests of
the SO(4) symmetry. We propose that the $\Theta$ excitation should
be observed as a sharp resonance in spin polarized inelastic
neutron scattering experiments in the superconducting phase. We
predict that the energy of the peak decreases toward the first
order phase transition to AF order. Such softening of modes is not
expected in general near a first order transition and would be a
unique signature of the enhanced symmetry at the transition point.

We thank S. Brown, P. Chaikin, M. Dressel, B.I. Halperin, C.
Kilic, S. Sachdev, A. Turner, D.-W. Wang, and S.C. Zhang for
useful discussions. This work was supported by Harvard NSEC.

\appendix

\section{Derivation of SO(3)$\times$SO(4) symmetry in Luttinger liquids}
\label{Bosonization}

In this Appendix we demonstrate that along the line $g_1=2g_2$,
the Luttinger liquid Hamiltonian (\ref{Luttinger_Hamiltonian}) has
an exact SO(3)$_{\rm spin}\times$SO(4)$_{\rm isospin}$ symmetry.
For this, we use bosonization to write
the $\Theta_\pm$ operators (\ref{ThetaOperator}) as ($r=\pm$)
\begin{eqnarray}
\Theta_r^\dagger=r\int dx\frac{\eta_{r\uparrow}\eta_{r\downarrow}}{2\pi\alpha}
e^{-rA(x)},
\label{bosonTheta}
\end{eqnarray}
where $A(x)=i\sqrt{2}\left(\phi_{\rho}(x)+\theta_{\rho}(x)\right)$
and $\theta_\rho=\pi\int^x dx'\Pi_\rho(x')$.
Note that $\Theta_r^\dagger$ are
independent of the spin fields $\phi_\sigma$ and $\theta_\sigma$.
Hence, the spin sector of the bosonized Hamiltonian commutes
trivially with $\Theta_r^\dagger$,
and we need only keep track of the charge sector,
\begin{eqnarray}
{\cal H}_\rho=\int dx\left(\frac{\pi u_\rho K_\rho}{2}\Pi_\rho^2+
\frac{u_\rho}{2\pi K_\rho}(\partial_x\phi_\rho)^2 \right).
\label{rhoHam}
\end{eqnarray}
Whenever $g_1=2g_2$, corresponding to $K_\rho=1$, the commutator
$[{\cal H}_\rho,A]$ takes on a simple form,
\begin{eqnarray}
\left[{\cal H}_\rho,A(x)\right]&=&\sqrt{2}u_\rho\left(\partial_x\phi_\rho(x)+\pi\Pi_\rho(x)\right)\nonumber\\
&=&-iu_\rho\partial_x A(x),
\end{eqnarray}
so that commuting ${\cal H}_\rho$ with an arbitrary function of $A(x)$ is equivalent to
taking the derivative with respect to $x$.  For example,
\begin{eqnarray}
\left[{\cal H}_\rho,e^{A}\right]&=&\sum_n \frac{1}{n!}\left[{\cal H}_\rho,A^n\right]\nonumber\\
&=&\left[{\cal H}_\rho,A\right]+\frac{1}{2}\left(A\left[{\cal H}_\rho,A\right]+
\left[{\cal H}_\rho,A\right]A(x)\right)+\ldots\nonumber\\
&=&-i u_\rho\left(\partial_x A+\frac{1}{2}\left( A\partial_x A+\partial_x A A\right)\right)
+\ldots\nonumber\\
&=&-i u_\rho\partial_x e^{A}.
\label{bosonCommutes}
\end{eqnarray}
Hence,
\begin{eqnarray}
\left[{\cal H}_\rho,\int dx\,e^{A(x)}\right]&=&-i u_\rho\int dx \,\partial_x e^{A(x)}\nonumber\\
&=&-i u_\rho \left[e^{A(L)}-e^{A(0)}\right],
\end{eqnarray}
which vanishes if periodic boundary conditions are imposed on $\phi(x)$ and $\theta(x)$.
Thus, for $K_\rho=1$,
\begin{eqnarray}
\left[{\cal H},\Theta_\pm^\dagger\right]=0,
\end{eqnarray}
and the Luttinger liquid has full SO(3)$_{\rm
spin}\times$SO(4)$_{\rm isospin}$ symmetry, generated by
$\Theta_\pm$, the total spin operators $S_\a$, and the charge of
left and right movers,
\be
Q_\pm=\sum_{ks} (a_{\pm,ks}^\dagger
a_{\pm,ks}-\frac{1}{2})
\ee
The enlarged symmetry relies on the
independent conservation of total number of right- and
left-movers.  This is not a good conservation law, for instance,
in the presence of impurity scattering, dimerization, or umklapp.
For general $K_\rho$, we find
\begin{eqnarray}
\left[{\cal H},\Theta_\pm^\dagger\right]&=&\frac{K_\rho^2-1}{2K_\rho}
\int dx\left(\sqrt{2}u_\rho(\partial_x \phi_\rho(x)-\pi\Pi_\rho(x))\right)
\frac{\eta_{\pm\uparrow}\eta_{\pm\downarrow}}{2\pi\alpha}
e^{\mp A(x)}.
\end{eqnarray}

We would like to thank Daw-Wei Wang for helpful discussions on results presented in this
section.

\section{ Parameters of the Ginzburg-Landau
free energy for weak interactions}
\label{AppendixGLDerivation}

\begin{figure}
\includegraphics[width=4cm]{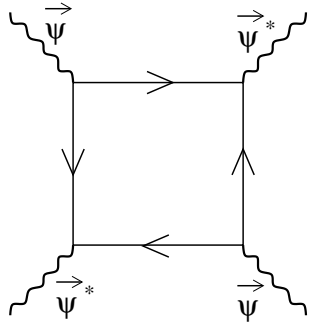}
\caption{Diagram for the coefficient $A$ in the GL free energy (\ref{effGL}).
\label{SCloop}}
\end{figure}

To extract parameters of the GL free energy we consider
a mean-field Hamiltonian
\begin{eqnarray}
{\cal H} &=& \sum_{ks}
(\epsilon_k - \mu) a_{ks}^\dagger a_{ks}
+\vec{\Psi}\cdot \sum_k {\rm w}_k a_{ks}^\dagger a_{-ks'}^\dagger
(\vec{\sigma}\sigma_2)_{ss'}
+\vec{\Psi}^*\cdot \sum_k {\rm w}_k a_{-ks'} a_{ks}
(\sigma_2\vec{\sigma})_{s' s}
\nonumber\\
&+&
\vec{\Phi}\cdot \sum_k a_{k-k_f s}^\dagger
a_{k+k_f s'} \vec{\sigma}_{ss'}
+\vec{\Phi^*}\cdot \sum_k a_{k+k_fs}^\dagger
a_{k-k_f s'} \vec{\sigma}_{ss'}
\end{eqnarray}
where w$_k=|k|/k$ gives the sign of $k$.
Integrating out the fermions yields an effective action
for the order parameter fields. We obtain the fourth order
terms:
\begin{eqnarray}
F_4= A \left( 2(|\vec{\Psi}|^2)^2 -
       |\vec{\Psi}^2|^2 \right)
+ B \left( 2(|\vec{\Phi}|^2)^2 -
       |\vec{\Phi}^2|^2 \right)
+ 2C|\vec{\Psi}|^2|\vec{\Phi}|^2
+ 2D \left( |\vec{\Psi}|^2|\vec{\Phi}|^2-
 |\vec{\Phi}\cdot \vec{\Psi}|^2-|\vec{\Phi}^*\cdot \vec{\Psi}|^2\right)
\label{effGL}
\end{eqnarray}
where
\begin{eqnarray}
A &=& \frac{1}{2\beta} \sum_{\omega_n} \int \frac{dk}{2\pi}
G^2(-k,-\omega_n) G^2(k,\omega_n)
\nonumber\\
B &=& \frac{1}{\beta} \sum_{\omega_n} \int \frac{dk}{2\pi}
G^2(k,\omega_n) G(k+2k_f,\omega_n)\left\{
G(k+2k_f,\omega_n)+2G(k-2k_f,\omega_n)
\right\}
\nonumber\\
C &=&-\frac{1}{\beta} \sum_{\omega_n} \int \frac{dk}{2\pi}
G^2(k,\omega_n) G(-k,-\omega_n)\left\{
G(k+2k_f,\omega_n)+G(k-2k_f,\omega_n)
\right\}
\nonumber\\
D &=&\frac{1}{\beta} \sum_{\omega_n} \int \frac{dk}{2\pi}
G(k,\omega_n)G(-k,-\omega_n)G(k+2k_f,\omega_n)G(-k-2k_f,-\omega_n)
{\rm w}_{-k}{\rm w}_{k+2k_f} \label{MFGLCoeff}
\end{eqnarray}
For instance, the diagram giving the coefficient $A$ is shown in
Fig. \ref{SCloop}.
For the Luttinger liquid type model with
linearized spectrum around $k=\pm k_f$
we obtain
\begin{eqnarray}
A&=&B=C/2=D. \label{GLisSO4sym}
\end{eqnarray}
The relationship among coefficients (\ref{GLisSO4sym})
implies that the effective GL
free energy (\ref{effGL}) is SO(3)$_{\rm spin}\times$SO(4)$_{\rm isospin}$
symmetric, as expected from the discussion in Section \ref{LuttingerSection}.
$F$ can thus be parameterized in the form (\ref{GinvGL}),
with $\tilde{u}_1=3A$ and $\tilde{u}_2=-2A$.  In the clean limit,
\begin{eqnarray}
A=\frac{7\zeta(3)}{16\pi^3v_f T^2}\nonumber
\end{eqnarray}
where $\zeta(3)=1.202\ldots$, and $v_f$ is the Fermi velocity.

\begin{figure}
\includegraphics[width=4cm]{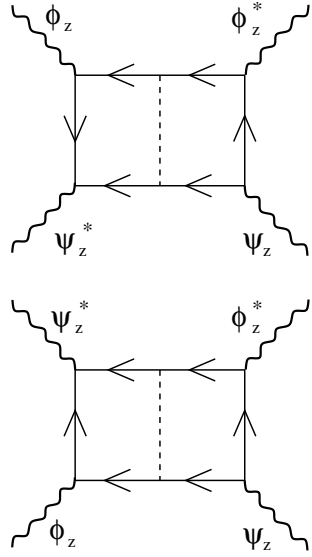}
\caption{Corrections to coefficient of
$|\Phi_z|^2|\Psi_z|^2$ term in GL free energy, to linear order in umklapp
scattering (dashed line).
The two diagrams add up to zero.  Inspection of all such diagrams shows
that the quartic coefficients of the GL free energy are not modified
to linear order.
\label{umklappCorrections}}
\end{figure}

Note that, as was pointed out in section
\ref{SectionHalfFillingGL}, to linear order in $g_3$, umklapp does
not affect the quartic coefficients of the free energy. For
instance, the diagrams in Fig. \ref{umklappCorrections} could
contribute to the coefficient of the term $|\Phi_z|^2|\Psi_z|^2$.
However, although they do not vanish individually, the two add up
to zero.  This is consistent with the Feynman-Hellman theorem
which requires that the only corrections to the free energy to
linear order in $g_3$ be given by the expectation value of the
perturbation (\ref{umkSDW}).

\section{Non-unitary triplet superconductivity
and antiferromagnetism. Expansion from the
lower critical dimension}

\label{AppendixB}

Here we outline some of the methods used in the RG calculation of the
non-linear model (\ref{FTriadContinuousLargeN}) in $d=2+\epsilon$
dimensions.
The RG flow equations of a general non-linear model can be computed using the
formalism of Friedan\cite{Friedan1985}.  The fields in such models must satisfy
constrains which force them to lie on some target space manifold ${\cal M}$.
For instance, the usual non-linear
$\sigma$ model deals with a single $N$-component vector with a constrained fixed length, and
${\cal M}$ in this case is the $N-1$ dimensional sphere describing the
locus of possible values of such vector.
A local set of coordinates $\phi_i$
can be introduced on a patch of ${\cal M}$,
in terms of which the free energy becomes
\begin{eqnarray}
F=\int dx\, g_{ij}(\phi(x))\partial_\mu\phi^i(x)\partial_\mu\phi^j(x).
\end{eqnarray}
Unlike the original fields used to define the model, the fields $\phi_i(x)$ are
unconstrained; all information regarding the original constrains is
contained in the metric $g_{ij}$.
The metric also contains the coupling constants
of the system.  In Friedan's formalism, the RG flow is thought of as
a gradual deformation of the manifold as the short length degrees of freedom
are integrated out. The RG equations can be written
in a covariant way; to one loop order,
\begin{eqnarray}
\frac{\partial g_{ij}}{\partial l}=\epsilon g_{ij}-R_{ij},
\end{eqnarray}
where $\epsilon=d-2$ and $R_{ij}$ is the Ricci tensor, which is determined uniquely by the
metric.

In practice, whenever the manifold is a homogeneous space $G/H$, as in our
case, it is simplest
to work directly in the tangent space of the manifold, see
\cite{Azaria1990} for a detailed discussion.
In terms of the metric on the tangent space, $\eta_{ab},$
the RG equations become
\begin{eqnarray}
\frac{\partial \eta_{ab}}{\partial l}=\epsilon \eta_{ab}-R_{ab},
\label{vielbeinRG}
\end{eqnarray}
where the Ricci tensor in the tangent space is given by
\begin{eqnarray}
R_{ab}=\sum_{Ic}f_{ac}\,^{I}f_{Ib}\,^{c}+\sum_{cd}
\frac{\eta_{aa}^2-(\eta_{cc}-\eta_{dd})^2}{4\eta_{cc}\eta_{dd}}
f_{ac}\,^{d}f_{bc}\,^{d}
\label{ricci}
\end{eqnarray}
in terms of the structure factor constants of the group $G$
\begin{eqnarray}
\left[T_a,T_b\right]&=&f_{ab}\,^{c}T_{c}+f_{ab}\,^{I}T_{I} \nonumber \\
\left[T_I,T_b\right]&=&f_{Ib}\,^c T_c.
\label{structFact}
\end{eqnarray}
Generators labelled by upper case indices are elements of Lie $H$,
while lower case indices denote generators in Lie $G$-Lie $H$.
In applying expression (\ref{ricci}), we assume that the generators
have been chosen so that the structure factor constants
(\ref{structFact}) are antisymmetric with respect to
exchange of any two indices; such a choice is always possible.
Equation (\ref{ricci}) is written in a non-covariant way
to make the dependence on the coupling
constants $\eta_{ab}$ explicit, and it shows the
advantage of working in tangent space: the Ricci tensor
is given directly in terms of the Lie algebra of $G$.

We briefly digress to discuss
the Lie algebra of the group $SO(N)$, which has
$N(N-1)/2$ generators corresponding to infinitesimal
rotations in the planes $\langle m,m' \rangle$, where the indices
$m\ne m'$ run through the $N$ independent axes.  For instance,
$SO(3)$ has three generators, $T_x=\langle \hat{y},\hat{z} \rangle$,
$T_y=\langle\hat{z},\hat{x}\rangle$, and
$T_z=\langle\hat{x},\hat{y}\rangle$.  Keeping in mind that $\langle m,m' \rangle=-\langle m',m \rangle$
(``a clockwise rotation in the $x-y$ plane is a counterclockwise
rotation in the $y-x$ plane"), we introduce a graphical
representation: if we draw $N$ points on a sheet of paper,
an arbitrary generator $\langle m,m' \rangle$ can be represented by an arrow
going from point $m$ to point $m'$.  The structure factor constants of the Lie algebra
\begin{eqnarray}
\left[\langle m,m'\rangle,\langle n,n'\rangle\right]=
\delta_{m'n}\langle m,n' \rangle-
\delta_{m'n'}\langle m,n \rangle-
\delta_{mn}\langle m',n' \rangle+
\delta_{mn'}\langle m',n \rangle \nonumber
\end{eqnarray}
can be written as a ``generalized
$\epsilon$ tensor",
\begin{eqnarray}
\left[\langle m,m' \rangle,\langle n,n' \rangle\right]
=\epsilon_{\langle m,m' \rangle\langle n,n'\rangle}\,^{\langle p,p' \rangle}
\langle p,p' \rangle,
\end{eqnarray}
which has a simple interpretation in terms of the arrows
described above: $\epsilon$ vanishes
unless $\langle m,m' \rangle$, $\langle n,n' \rangle$, and $\langle p,p' \rangle$ are
the edges of a closed triangle.  If they do form a closed
triangle, count the number of times that the directions of
the arrows must be flipped to turn it into an oriented
triangle, {\it i.e.} one satisfying $m'=n$, $n'=p$, and $p'=m$.
If the number of flips is even, then $\epsilon=1$; otherwise
$\epsilon=-1$.  With this in mind, inspection of equation
(\ref{ricci}) shows that, for groups $G$ based on $SO(N)$,
where $f_{ab}\,^c\propto \epsilon_{ab}\,^c$,
the calculation of the Ricci tensor
reduces almost entirely to the counting of triangles.

Armed with these tools, consider the non-linear model (\ref{FTriadContinuousLargeN}),
\begin{eqnarray}
F = \int d^d x \left\{ \frac{1}{2g_1} (\nabla \vec{e}_1)^2+
\frac{1}{2g_2}((\nabla \vec{e}_2)^2+(\nabla \vec{e}_3)^2)
+\frac{1}{g_3}((\vec{e}_1 \cdot \nabla \vec{e}_2)^2 +(\vec{e}_1 \cdot \nabla
\vec{e}_3)^2) +\frac{1}{g_4} (\vec{e}_2\cdot \nabla \vec{e}_3)^2
\right\}.
\label{vecModel}
\end{eqnarray}
Model (\ref{vecModel}) has the symmetry $SO(N)$ of rotations
of the $N$-component vectors, and the symmetry $SO(2)$
of internal rotations between $\vec{e}_2$ and $\vec{e}_3$.  Hence,
the symmetry group of (\ref{vecModel}) is $G=SO(N)\times SO(2)$.
The order parameter is a triad of mutually orthogonal vectors,
$\Phi=(\vec{e}_1\,\,\vec{e}_2\,\,\vec{e}_3)$, and the ordered phase
has residual symmetry $H=SO(N-3)\times SO(2)_{\rm diag}$.
The generators of $H$ leave the triad $\Phi$ invariant, whereas the
generators in Lie $G$-Lie $H$ rotate the triad and
are in one-to-one correspondence with the spin waves of the system.

\begin{figure}
\includegraphics[width=12cm]{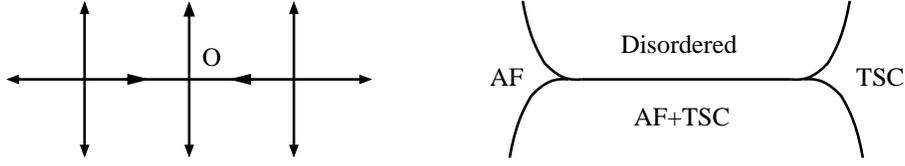}
\caption{RG flows in the equations (\ref{RG2+eLargeN}) for $N=4$.
Point O has a symmetry $SO(4)\times SO(4)$. \label{FigLargeN2E}}
\end{figure}

We identify four types of spin waves, corresponding to the
following classes of generators: $T_{a_1}$, which leave
$\vec{e}_{\{2,3\}}$ untouched but
rotate $\vec{e}_1$ into one of the remaining $N-3$
directions; $T_{a_2}$, which leave $\vec{e}_1$ untouched,
but rotate either $\vec{e}_2$ or $\vec{e}_3$ into one of the
remaining $N-3$ directions;  $T_{a_3}$, of rotations in either
the $\vec{e}_1,\vec{e}_2$ plane or the  $\vec{e}_1,\vec{e}_3$
plane; and $T_{a_4}$, composed of the single generator of
rotations in the $\vec{e}_2,\vec{e}_3$ plane.
Each class furnishes an independent irreducible representation under the
action of the group H, leading to four different spin wave velocities,
and to four different coupling constants, $\eta_1\ldots\eta_4$,
\begin{eqnarray}
\eta_{bc}=\sum_{a_1}\eta_1 \delta_{b a_1}\delta_{c a_1}+
\sum_{a_2}\eta_2 \delta_{b a_2}\delta_{c a_2}+
\sum_{a_3}\eta_3 \delta_{b a_3}\delta_{c a_3}+
\eta_4 \delta_{b a_4}\delta_{c a_4}.\nonumber
\end{eqnarray}
The RG flow equations (\ref{vielbeinRG}) become
\begin{eqnarray}
\frac{d\eta_1}{dl} &=& \epsilon \eta_1 -
\frac{1}{2\pi}\left(N-2+\frac{\eta_1^2-\eta_2^2-\eta_3^2}{\eta_2\eta_3}\right)
\nonumber\\
\frac{d\eta_2}{dl} &=& \epsilon \eta_2 -
\frac{1}{2\pi}\left(N-2+\frac{\eta_2^2-\eta_3^2-\eta_1^2}{2\eta_1\eta_3}
-\frac{\eta_4}{2\eta_2}\right)
\nonumber\\
\frac{d\eta_3}{dl} &=& \epsilon \eta_3 -
\frac{1}{2\pi}\left(N-2+\frac{N-3}{2}\frac{\eta_3^2-\eta_1^2-\eta_2^2}{\eta_1\eta_2}
-\frac{\eta_4}{2\eta_3}\right)
\nonumber\\
\frac{d\eta_4}{dl} &=& \epsilon \eta_4 -
\frac{1}{2\pi}\left(\frac{N-3}{2}\frac{\eta_4^2}{\eta_2^2}+\frac{\eta_4^2}{2\eta_3^2}\right)
\label{AppCresult}
\end{eqnarray}
The fixed $SO(N)\times SO(3)$ symmetric point of (\ref{AppCresult}) is described in
the body of the text for $N\ge 5$. For $N=4$, the fixed point is stable with respect to arbitrary
perturbations away from the $SO(N)\times SO(3)$ symmetric plane [incidentally, in this
case the fixed point has a larger symmetry than expected,
$SO(4)\times SO(4)$].  On the other hand, within the plane,
it has one stable and one unstable direction.  This suggests the RG flows and the phase
diagram that are shown schematically in Fig. \ref{FigLargeN2E}.
Note that we have a whole  line of direct transitions between the
disordered and the TSC+AF phases. This whole line is controlled by
the point  O that has a high $SO(4)\times SO(4)$ symmetry. This is
quite remarkable: a higher symmetry appears not at a single point
but at the whole transition line.

\section{Non-unitary triplet superconductivity
and antiferromagnetism. Large $N$ analysis}

\label{AppendixC}

In Section \ref{SectionNonUnitaryFiniteTLargeN} we pointed out that a free
energy in (\ref{NearlySO4GLenergy}) in the large $N$ limit should be
considered with care, when $\tilde{u}_1+\tilde{u}_2/3$ is close to
zero.  Here we assume that $\tilde{u}_2>0$, so, this requires
negative $\tilde{u}_1$. The complications arise when the system goes
outside the basin of attraction of the tetracritical fixed point, and
the RG flows carry $\tilde{u}_1$ to large negative values. As we discuss
below, this leads to a first order transition which is similar to what
was suggested in Ref. \onlinecite{Bailin1977} for the normal to
$\mathrm{A}_1$ transition in liquid $^3\mathrm{He}$.
We take
\begin{equation}
\tilde{u}_1+\tilde{u}_2/3=\frac{\delta}{4N}
\label{eq.cond_small_illdef}
\end{equation}
where $\delta$ is positive and is of order $\frac{1}{N}$.
We now extend the calculations
presented in Section \ref{SectionNonUnitaryFiniteTLargeN}
to the next order in $1/N$.
For all order parameters we separate  expectation values and
fluctuations
\begin{eqnarray}
\vec{\Psi}&=&(\vec{a}_T+\imath\vec{b}_T,\sigma_\psi+a_L+
\imath b_1,a_0+\imath\sigma_\psi+\imath b_L,a_1+\imath b_0),
\nonumber\\
\vec{N}&=&(\vec{N}_T,N_0,N_1,\sigma_N+N_L),
\label{ExpansionAppendixC}
\end{eqnarray}
We can expand equation (\ref{NearlySO4GLenergy})
to order $1/N^2$ and obtain tadpole equations for $a_L$ and $N_L$.
In addition to the counterterms and
loops due to fluctuations of the transverse components,
we need to include fluctuations of the longitudinal
components. Note, that loops of longitudinal
components may be terminated by bubble chains coming
from $\tilde{u}_2(\vec{a}_T\vec{b}_T)a_0$
vertices. We also need
to include diagrams that arise from
$\tilde{u}_2(\vec{a}_T\vec{b}_T)b_1a_L$
vertices. Special attention
should be paid to diagonalization of propagators,
since the free energy has terms which introduce
mixing between fluctuating components
in (\ref{ExpansionAppendixC}).

If we want to absorb the cut-off dependence into renormalization of
quadratic coefficient (compare to equation (\ref{rcdefinition})), we
need to define the latter relative to
\begin{equation}
r_{c'}=r_c-(40\tilde{u}_1+24\tilde{u}_2)\int_0^\Lambda
\frac{d^3k}{(2\pi)^3}\frac{1}{k^2}+
(4\tilde{u}_1+8\tilde{u}_2)\frac{j}{2\pi^2}\, \log \Lambda,
\end{equation}
where
\begin{equation}
j=\frac{\tilde{u}_2N}{4}.
\label{eq.j}
\end{equation}

Integrals in tadpole equations can not be calculated exactly. Hence,
we expand them in two cases: $j^2\gg
32\tilde{u}_2\sigma^2$ and $j^2\ll 32\tilde{u}_2\sigma^2$ ($\sigma^2$
corresponds to $\sigma^2_N$ or $\sigma^2_\Psi$ or
$(\sigma^2_N+\sigma^2_\Psi)/2$ depending on terms in the integrals).
Also, while solving final system of equations, expansions under
conditions $\sigma^2_\Psi\gg\sigma^2_N$, $\sigma^2_\Psi\ll\sigma^2_N$,
$\sigma^2_\Psi\approx\sigma^2_N$ were made.  To be concrete, we took
the values $\tilde{u}_1=-1/(4N)+\delta/(4N)$ and $\tilde{u}_2=3/(4N)$.

Transition from disordered phase to superconductive and antiferromagnetic phases
outside the vicinity of $r_{c'}$ remains of the second order, though transition
border shifts such that
\begin{equation}
t_{N,\Psi}=\frac{15}{16 \pi^2N}\log\frac{16}{3}.
\end{equation}

In the vicinity of full $SO(3)\times SO(N)$ symmetry line
$t_N=t_\Psi$ we expanded equations under conditions
\begin{equation}
|\sigma^2_N-\sigma^2_\Psi|\sim\sigma^2/N,
\label{eq.cond_small_sigm_diff}
\end{equation}
resulting in
first order phase transition, limited by boundaries
\begin{equation}
t_{N,L}+2t_{\Psi,L}=0
\label{eq.tL_nonun_N}
\end{equation}
and
\begin{equation}
t_{N,M}+2t_{\Psi,M}=\frac{C^2_0}{4N^2\delta},
\label{eq.tM_nonun_N}
\end{equation}
where $C_0=(1+3\sqrt{3/2})/\pi$.

\begin{figure}
\includegraphics[height=5cm]{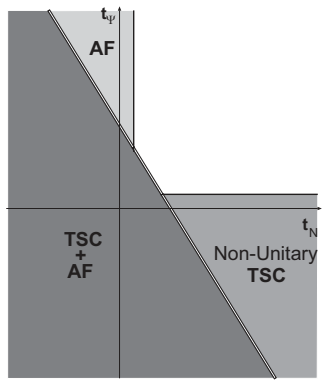}
\caption{Phase diagram for large N under
conditions~(\ref{eq.cond_small_illdef}).  Solid line is a I order
phase transition, dashed -- II.
\label{fig.ptrans_N_cond}}
\end{figure}

Condition~(\ref{eq.cond_small_sigm_diff}) for solution obtained,
appears to be valid not only for small deviations
from the line of symmetry, but for entire line of transition.
Thus solution~(\ref{eq.tL_nonun_N},\ref{eq.tM_nonun_N}) is self-consistent
in the entire region, resulting in phase diagram shown on
Fig.~\ref{fig.ptrans_N_cond}.
In comparison with solution of the first order expansion,
boundary of mixed phase becomes a first order phase transition,
and there is no angle between $N\Psi\leftrightarrow N$ and
$N\Psi\leftrightarrow \Psi$ boundaries
(which is $\propto\delta$ in first order expansion).
Boundary of the basin of attraction of stable fixed point
is determined by the validity
of expansion for different conditions for $j^2$.
In our case  it is $N\delta\sim 1$.

On the other hand, expansion under condition $\sigma^2_N\gg\sigma^2_\Psi$
for $N\Psi\leftrightarrow N$ transition,
and $\sigma^2_N\ll\sigma^2_\Psi$ for $N\Psi\leftrightarrow \Psi$
also results in self-consistent solution
with phase transition boundary of a different geometry. On the
boundary minor component drops to zero, while
major almost does not change. In our opinion, this solution
does not have
a physical sense.


\begin{thebibliography}{124}
\bibitem[{end({\natexlab{b}})}]{endnoteTSC}
\bibinfo{note}{{Triplet pairing in Bechgaard salts does not have orbital degree
  of freedom. Quasi one-dimensional electron band structure requires
  $\vec{\Psi}(\vec{p}) \propto p_x$, where $x$ is the direction parallel to the
  chains. This provides considerable simplification in writing effective
  models, such as the Ginzburg-Landau free energy (\ref{GLenergy})}}.
\end{thebibliography}
\end{document}